\documentclass[aps,prapplied,superscriptaddress,twocolumn,floatfix,showpacs]{revtex4-2}
\usepackage{graphicx,color}
\usepackage[utf8]{inputenc}
\usepackage{amsmath,,amssymb,bm,amsfonts,dsfont,mathrsfs,amsthm}
\usepackage{braket}
\usepackage{txfonts,comment}
\usepackage[colorlinks=true,linkcolor=blue,citecolor=blue,urlcolor=blue]{hyperref}
\usepackage{soul}
\DeclareMathOperator{\sech}{sech}

\begin{document}

\title{Saturable Global Quantum Sensing}

\author{Chiranjib Mukhopadhyay} 
\affiliation{Institute of Fundamental and Frontier Sciences, University of Electronic Sciences and Technology of China, Chengdu 611731, China}
\affiliation{Key Laboratory of Quantum Physics and Photonic Quantum Information, Ministry of Education,
University of Electronic Science and Technology of China, Chengdu 611731, China}

\author{Matteo G. A. Paris}
\affiliation{Dipartimento di Fisica, Università di Milano, I-20133 Milan, Italy}

\author{Abolfazl Bayat}
\affiliation{Institute of Fundamental and Frontier Sciences, University of Electronic Sciences and Technology of China, Chengdu 611731, China}
\affiliation{Key Laboratory of Quantum Physics and Photonic Quantum Information, Ministry of Education,
University of Electronic Science and Technology of China, Chengdu 611731, China}

\begin{abstract}
Conventional formulation of quantum sensing has been mostly developed in the context of local estimation, 
where the unknown parameter is roughly known. In contrast, global sensing, where the prior information is 
incomplete and the unknown parameter is only known to lie within a broad interval, is practically more 
engaging but has received far less theoretical attention. Available formulations of global sensing rely 
on adaptive Bayesian strategies requiring on-the-fly change in measurement settings, or minimizing average uncertainty yielding unsaturable bounds. Here, we provide an operationally motivated approach 
to global sensing for fixed but optimized settings. Our scheme yields a saturable precision bound optimizing the measurement as well as the probe preparation simultaneously. The formalism is general and computationally scalable for generic bosonic multimode Gaussian or many-particle free-fermionic quantum sensors. We illustrate the implications for Gaussian thermometry and Gaussian phase estimation by showing that the optimal measurement changes, either gradually or abruptly, from homodyne for local sensing, towards heterodyne for 
global sensing.In contrast, for fermionic transverse XY probes, the optimal measurement basis stays fixed independent of width. 
\end{abstract}

\maketitle

\section{Introduction} Quantum metrology~\cite{giovannetti2004quantum,giovannetti2006quantum,giovannetti2011advances,toth2014quantum,Paris_2016,qvarfort2018gravimetry, montenegro2024quantum,he2023stark,free2022sarkar, de2018genuine,de2018quantum,correa2015individual,mehboudi2019thermometry,mukhopadhyay2018superposition,pati2020quantum,zhao2020quantum,aslam2023quantum, ye2024essay,yu2024experimental,gribben2024quantum} is underpinned by the quantum theory of local estimation via the celebrated Cram\'er-Rao (CR) theorem~\cite{cramer1999mathematical,rao1992information,helstrom1969quantum,statistical1994braunstein}, 
which sets the ultimate limits on the precision of any detector. The term \emph{local} means the parameter to be sensed is roughly known, i.e., we have significant prior information about 
its value already~\cite{paris2009quantum,braun2018quantum,degen2017quantum,liu2019quantum, montenegro2024quantum, mukhopadhyay2024modular}. This prior information thus allows us to optimize a measurement setting for obtaining the best possible precision. In contrast, in practical situations prior information is typically incomplete, reflecting our ignorance of the unknown parameter. Thus, a formulation of \emph{global} quantum estimation with incomplete prior information yielding an optimal measurement strategy is paramount for practical use of quantum sensors. 

Unlike local sensing which is fully characterized by CR theorem, there is less agreement on a quantitative figure of merit for global estimation tasks \cite{rubio2019quantum,rubio2021global,rubio2022quantum,mok2021optimal,zhou2024strict,mihailescu2024uncertain}.
\textcolor{black}{
In general, some cost function should be introduced to assess 
the estimate and then the optimal strategy minimizes the average cost \cite{helstrom1969quantum}. Few solutions to this problem are known, notably for the estimation of a shift-parameter \cite{d1998general}.
Alternatively, one may introduce a broader prior and derive the so-called Van Trees inequality for the averaged variance \cite{vantrees1968}.}
In a Bayesian approach, one adaptively 
updates the measurement basis at each round, reducing the global sensing problem into local sensing iteratively~\cite{jarzyna2015true,rubio2020bayesian,cimini2024benchmarking,salvia2023critical}, where the 
sensing precision has been formulated through Van Trees inequalities~\cite{van2004detection,gill1995applications}. 
To avoid potentially challenging adaptive measurements, an alternative figure of merit for global sensing has been put forward in Ref.~\cite{montenegro2021global}, which introduces a bound on the average estimation uncertainty over the entire interval that the unknown parameter is assumed to be located in. However, none of these two previous methods help find a saturable bound for a single fixed strategy. Thus, a natural question is to ask whether it is possible to find a saturable formulation for global sensing in this setting. 

In this paper, we answer this question affirmatively by formulating an alternative approach for global sensing, which bounds the average uncertainty with a quantity saturable by construction. Our strategy can be used to optimize over both probe preparation as well as measurement parameters. We demonstrate the implications for two paradigmatic classes of quantum sensors, viz., optical Gaussian metrology \cite{serafini2017quantum,ferraro2005gaussian,wang2007quantum,weedbrook2012gaussian,adesso2014continuous,serafini2003symplectic,giedke2002characterization,filip2013distillation, serafini2017quantum,rugar1991mechanical,olivares2007optimized,sparaciari2015bounds,adesso2014gaussian,sanz2017quantum,friis2015heisenberg,ruppert2017light,matsubara2019optimal,motes2015linear,sparaciari2016gaussian,nichols2018multiparameter,huang2008optimized, yadin2018operational,laurenza2018channel, correa2015individual,oh2019optimalpra,oh2019optimal,bakmou2019quantum,sorelli2024gaussian,porto2024enhancing,liao2022quantum, hayashi2022global} {\color{black} and many-body fermionic ground-state Gaussian metrology} \cite{montenegro2024quantum,zanardi2006ground,ye2024essay}, where the corresponding theory of local sensing is highly developed, and experimental implementations are  underway ~\cite{pirandola2018advances,dowling2015quantum,oh2017practical,raffaelli2018homodyne,wang2019heisenberg,wolf2019motional,mccormick2019quantum, pan2024realisation}. 

\section{Quantum estimation theory} The goal of any quantum sensing protocol is to estimate an unknown parameter $\lambda$, encoded in the quantum state $\rho(\lambda)$, by performing a measurement described by Positive-Operator Valued Measure (POVM) 
operators $\Pi=\{ \Pi_k\}$ whose measurement outcomes  occur with probabilities $p_k(\lambda){=}\mathrm{Tr}\left[ \Pi_k \rho(\lambda) \right]$. The estimated value  $\lambda_{est}$ is then obtained  through  collecting and post-processing of $M$-rounds of observed experimental data. The precision of the estimation, quantified by variance of difference between $\lambda_{est}$ and the true value of parameter $\lambda$ , is then fundamentally bounded by the CR inequality~\cite{fisher1922mathematical,cramer1999mathematical,rao1992information}  
\begin{equation}
    \text{Var}[\lambda_{est} - \lambda ] \geq \frac{1}{M F_c (\lambda,\Pi)},
    \label{cramer-rao}
\end{equation}
where $F_c(\lambda,\Pi){=}\sum_k p_k(\lambda) [\partial_\lambda \ln p_k(\lambda)]^2$ is called the \emph{classical Fisher information} (CFI). The ultimate bound obtained via optimizing over all measurements $\Pi$, is given through the \emph{quantum Fisher information} (QFI) $F_Q{=}\max_{\Pi}F_c(\lambda, \Pi)$, and reads~\cite{paris2009quantum}
\begin{equation}
    \text{Var}[\lambda_{est} - \lambda ] \geq \frac{1}{M F_c(\lambda,\Pi) } \geq \frac{1}{M F_Q (\lambda)}.
    \label{cramer-rao_q}
\end{equation}
The optimal measurement is along the eigenbasis of operator $L_{\lambda}$ satisfying the implicit equation $\partial_{\lambda}\rho(\lambda) {=} [\rho(\lambda)L_{\lambda}{+} L_{\lambda} \rho(\lambda)]{/}2$. 

\section{Global sensing} 

The optimal measurement saturating the CR inequality Eq.~\eqref{cramer-rao_q} generally depends on the unknown parameter $\lambda$. Therefore, it can only be achieved for local sensing problem where $\lambda{\in} [\lambda_0{-}\Delta/2,\lambda_0{+}\Delta/2]$ varies over a small interval $\Delta {\rightarrow} 0$ for the optimal measurement strategy at $\lambda{=}\lambda_0$.  On the other hand, for global sensing where $\Delta$ is large, no single measurement basis is necessarily optimal for the entire sensing interval. In this situation, a \textcolor{black}{different approach} is needed to optimize the precision of sensing \textcolor{black}{and to assess metrological protocols}. Our initial knowledge of the unknown parameter $\lambda$ is encoded in a prior distribution $\mathcal{P}(\lambda)$.  The first formulation of global sensing has been developed by van Trees~\cite{van2004detection} in the context of classical estimation theory, later extended to the quantum setting~\cite{kolodynski2010phase,demkowicz2012elusive,martinez2017quantum}. However, achieving this bound requires Bayesian updating, thus earning the name \emph{Bayesian CR bound}~\cite{gill1995applications}. A different approach was developed in Ref.~\cite{montenegro2021global} in which the global average uncertainty per round  was introduced as a figure of merit 
\begin{align} \label{eq:Global_QFI}
\int_{\lambda_0{-}\Delta/2}^{\lambda_0{+}\Delta/2}
\!\!\!\!\!\! d\lambda'
\mathcal{P}(\lambda') \text{Var}[\lambda'_{est} - \lambda' ]  \ge & \int_{\lambda_0{-}\Delta/2}^{\lambda_0{+}\Delta/2} 
\!\!\!\!\!\! d\lambda'
\frac{\mathcal{P}(\lambda')}{ F_Q(\lambda')} \notag \\ \, & = \, \, G(\lambda_0, \Delta)
\end{align}
However, this new bound still cannot be saturated in the general case where the optimal measurement depends on $\lambda$. For a more complete discussion of global quantum estimation, the interested reader may wish to consult Ref.~\cite{mukhopadhyay2025current}.

\subsection{Saturable global uncertainty} 

In order to obtain a saturable bound, we start from the original CR inequality of Eq.~(\ref{cramer-rao}) in which the measurement setup $\Pi$ is fixed. One can define the
average global uncertainty with respect to the fixed measurement basis $\Pi$ by averaging over the both sides of Eq.~(\ref{cramer-rao}). A Saturable Global Uncertainty (SGU) can be obtained by minimizing such average global uncertainty with respect to all measurement bases
\begin{align}
      \int_{\lambda_0{-}\frac{\Delta}{2}}^{\lambda_0{+}\frac{\Delta}{2}} \!\!\!\!\!d\lambda'\, \mathcal{P}(\lambda') \text{Var}[\lambda'_{est} - \lambda' ]  {\geq} & \min_{\Pi}\int_{\lambda_0  {-} \frac{\Delta}{2}}^{\lambda_0 {+} 
      \frac{\Delta}{2}} \!\!\!\!\!  d\lambda' \frac{\mathcal{P}(\lambda')}{F_{c} \left(\lambda', \Pi \right)}  \notag \\ & \, {=} \,\,\mathcal{G}(\lambda_0, \Delta), 
    \label{eq:avg_uncertainty}
\end{align}

For a fixed $\Pi$ we use the notation $\tilde{\mathcal{G}} (\lambda_0, \Delta, \Pi)$, such that $\mathcal{G} {=} \min_{\Pi} \tilde{\mathcal{G}}$. Note that, the right hand side of the above inequality results in a single optimal measurement basis which is used for sensing over the entire interval without any dependence on the exact value of the unknown parameter $\lambda$. To reiterate, this figure of merit, the SGU, is the average of local variances weighted with requisite priors. For a fixed measurement setting, these local variances are individually lower bounded by Classical Fisher Information for that measurement setting, Thus, our optimal measurement, which we shall identify with various examples in the following sections, is the single fixed measurement setting that minimizes this SGU, or, in other words, minimizes the prior-weighted average of the inverse CFIs across the estimation interval. It is in this sense we identify the \emph{optimal} measurement. This ultimate bound is thus naturally saturable by construction. \textcolor{black}{For concreteness, we assume an uniform prior, however, other priors may be more suitable for specific scenarios, as for instance Jeffrie's prior has been used for scale-invariant sensing~\cite{rubio2021global}. } Moreover, when the probe preparation also depends on the uncertain value of a different nuisance parameter $\xi {\in} [\xi_{\min},\xi_{\max}]$, the SGU can be readily extended as $$\mathcal{G}{=}\min_{\Pi}\int_{\lambda_0-\frac{\Delta}{2}}^{\lambda_0+\frac{\Delta}{2}}\!\!\!\! d\lambda' \int_{\xi_{\min}}^{\xi_{\max}}\!\!\!\!d\xi'\,\frac{\mathcal{P}(\lambda') \mathcal{\tilde{P}}(\xi')}{F_{c}(\lambda',\xi',\Pi)}\,,$$ where $\tilde{\mathcal{P}}$ represents our prior knowledge about the distribution of loss/nuisance parameter $\xi$. Note that our bound can also be exploited in adaptive quantum sensing, where $\mathcal{G}$ identifies the optimal measurement setting for a prior distribution given by $\mathcal{P}(\lambda)$. After each round of sampling with this optimal measurement, one can update the prior distribution and repeat the procedure to specify the new optimal measurement basis to continue with the next round of sampling. \\

We explicitly note that Ref~\cite{zhou2024strict} and related references define a different figure of merit as a single global variance, namely
$\mathrm{Var}[\hat{\Lambda}] = \sum_{x} p(x|\lambda) [\hat{\Lambda}(x) - \lambda]^2$. This quantity represents the average squared error of a global estimator and admits, in principle, a saturable bound analogous to the van Trees inequality. Ref.~\cite{zhou2024strict} thereafter investigates several strategies -- parallel, sequential, and those involving indefinite causal order and establishes a strict hierarchy therein, in contrast with local estimation~\cite{giovannetti2006quantum}. It is worth emphasizing that parallel strategies, which are closest in spirit to our setting (i.e., no inter-round adaptative change), are demonstrated numerically~\cite{zhou2024strict} to be the least effective in achieving this saturation. In contrast, this work focuses on a different figure of merit, viz., the saturable global uncertainty (SGU), which is defined as a weighted sum over local variances. We emphasize that this is \emph{not} the equivalent to the global variance of Ref.~\cite{zhou2024strict}. Within the constraints of this definition of figure of merit, we derive and explicitly construct a saturable bound for SGU, demonstrating its operational value in realistic measurement settings. At this point, one may wonder whether the optimization for SGU vide Eq.~\eqref{eq:avg_uncertainty} is computationally feasible or becomes prohibitively complicated with increase in sensor size. For the rest of this work, we demonstrate with concrete examples that the SGU can indeed be efficiently computed for a wide range of sensing platforms.  

\section{Bosonic Gaussian global sensing}  Let us first discuss the photonic Gaussian sensing paradigm. The probe is fully described by a Gaussian quantum state $\rho (\lambda)$ completely characterized by the displacement vector $d$ and the covariance matrix $\sigma$. Likewise,  we assume  Gaussian measurements $\Pi$, also represented in terms of displacement $d_m$ and covariance matrix $\sigma_m$. The corresponding CFI is then given as~\cite{cenni2022thermometry}
\begin{equation}
    F_c = \frac{\partial d^T}{\partial \lambda} (\sigma {+} \sigma_m)^{-1}\frac{\partial d}{\partial \lambda} {+} \frac{1}{2} \textrm{Tr} \left[\left((\sigma {+} \sigma_m)^{-1} \frac{\partial \sigma}{\partial\lambda}\right)^2\right]
    \label{eq:fc_gaussian}
\end{equation} 
From this expression, it is enough to optimize over zero-displacement general-dyne measurements \cite{genoni2014general} characterized by diagonal covariance matrices $\sigma_m{=}\text{diag}(r_m, 1/r_m)$, i.e., along directions rotated at a generic angle from quadrature measurements. As two well-known limits of such general-dyne measurements, there exists homodyne measurements ($r_m {=}0$) along quadrature directions, and complementary heterodyne measurements ($r_m {=} 1$) along coherent state directions.

\subsection{Gaussian thermometry} We begin by considering the simplest case of equilibrium Gaussian thermometry for a single-mode system described by the Hamiltonian $H{=}\omega a^\dagger a$, where $a$ ($a^\dagger$) is the annihilation (creation) operator and energy scale $\omega{=}1$. The system is considered to be in a thermal state at an unknown temperature $T {=} 1/\beta$, with covariance matrix $\sigma {=} \nu \mathbb{I}$, where $\nu = \coth(\beta\omega/2)$. We optimize SGU over general-dyne measurements with respect to parameter $r_m$. The expression for CFI with respect to the temperature $T$ is given by 
\begin{equation}
    F_c = \frac{(\partial_T\nu)^2}{2} \left[ \frac{1}{(\nu + r_m)^2} + \frac{1}{(\nu + 1/r_m)^2} \right]
\end{equation}

\noindent Assuming a uniform prior within the known interval $T{\in}[T_{0} {-}\Delta/2,T_{0}{+}\Delta/2 ]$, the expression for SGU can now be put in the following simplified form
\begin{equation}
    \mathcal{G}(T_0, \Delta) = \min_{r_m}\frac{1}{\Delta}\int_{T_0 - \Delta/2}^{T_0 + \Delta/2}  \frac{2 d\nu}{(\partial_T\nu)^2 \left[ \frac{1}{(\nu + r_m)^2} + \frac{1}{(\nu + 1/r_m)^2} \right] }, 
    \label{avg_uncertainty_thermometry}
\end{equation}
\begin{figure}[h]
	\centering
	\includegraphics[width = 0.5\textwidth]{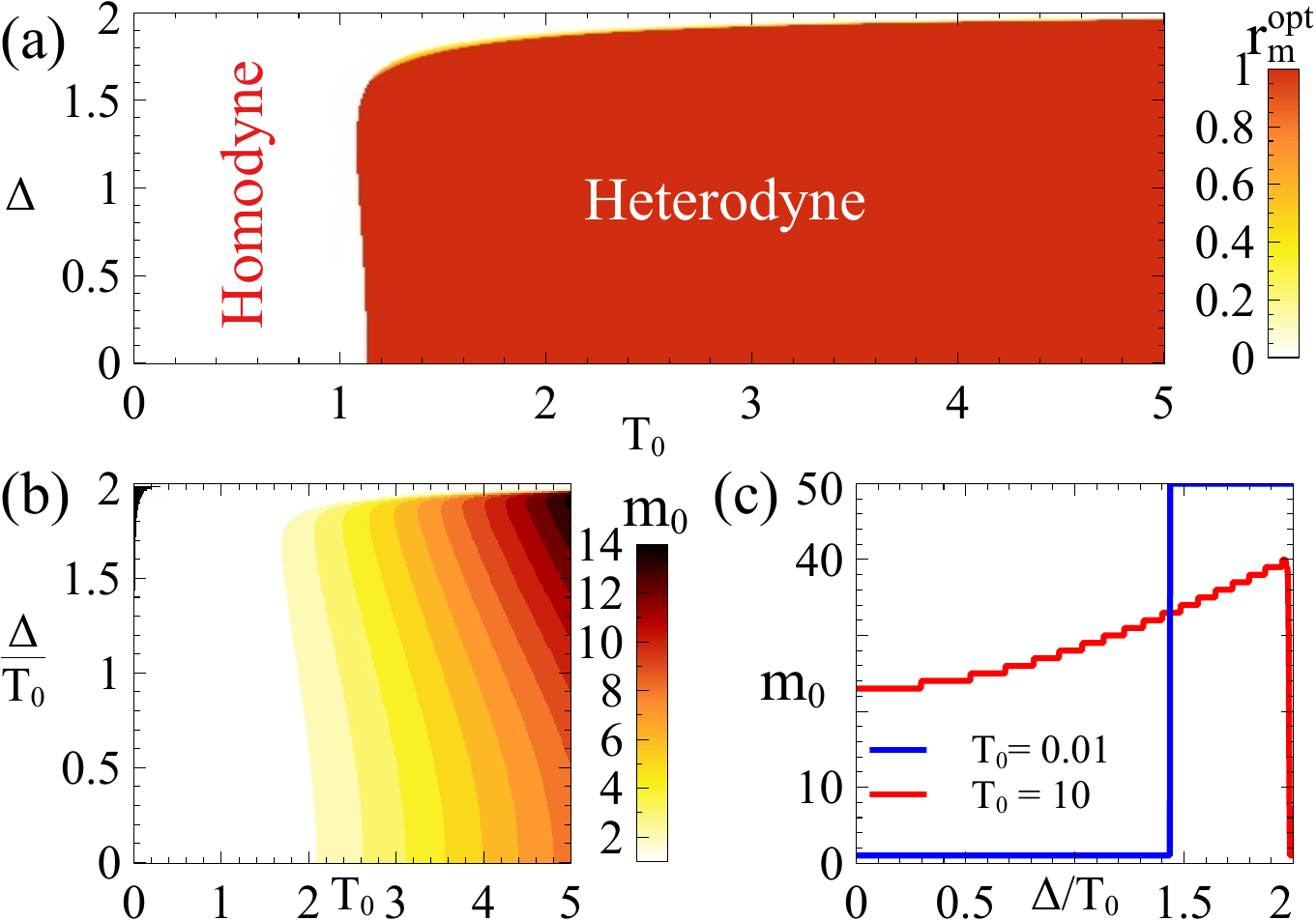}
	\caption{ (a) Density plot of optimal measurement strategy $r_m{=}r_m^{\text{opt}}$ for varying temperature $T_0$ and relative width $\Delta/T_0$. (b) Density plot of minimal levels $m_0$ required for photon counters to outperform optimal Gaussian measurements with $T_0$ and $\Delta/T_0$. (c) $m_0$ vs $\Delta/T_0$ for $T_0 = 0.01$ (blue) and $T_0 = 10$ (red). $\omega{=}1$, $\Delta/T_0 \leq 2$ throughout. }
	\label{fig:gaussian_thermometry}
\end{figure}
This integration is expressible in terms of elementary functions and the subsequent minimization with respect to the measurement squeezing parameter $r_m$ can be performed analytically via Mathematica. For local thermometry~\cite{cenni2022thermometry}, the optimal measurement strategy is either a homodyne (in the low temperature regime) or a heterodyne (in the high temperature regime) and a transition temperature $T_c {\approx} 1.13$ at which an abrupt flip occurs. Note that, although the measurement basis changes abruptly it should not be considered as a quantum phase transition. In fact, this abrupt change in the optimal measurement basis is due to the contribution of higher Fock states contributing more and more to the thermal state. Indeed, this is merely a single-mode harmonic oscillator and there is no competing terms in the Hamiltonian to induce a phase transition. Interestingly, in our global paradigm, the optimal measurement remains either a homodyne ($r_m{=}0$) or a heterodyne ($r_m{=}1$), as depicted in Fig.~\ref{fig:gaussian_thermometry}(a). However, the transition temperature at which the optimal measurement strategy flips from a homodyne to heterodyne is now dependent on the width of the temperature window $\Delta$. Of course, the truly optimal, albeit non-Gaussian, measurement setup for thermometry is the Fock basis, i.e., a perfect photon counter which can resolve each of the countably infinite levels. However, in practice photon counters can at best perfectly resolve up to a finite number of Fock levels. One may ask how many levels, say $m_0$, are required for outperforming the best possible Gaussian measurement. In Fig.~\ref{fig:gaussian_thermometry}(b), we plot $m_0$ as a function of $T_0$ and $\Delta/T_0$, and notice that for low temperatures and not too extreme widths, even a simple on-off photon detector ($m_0 {=} 1$) outperforms the optimal Gaussian measurement protocol. However, for higher temperatures, $m_0$ increases rapidly with both temperature and width, i.e., Gaussian heterodyne measurements become optimal. Notice that there is a small region around $T_{0}{\simeq}0$ and $\Delta/ T_0{{\simeq}} 2$, see Fig.~\ref{fig:gaussian_thermometry}(b), that $m_0$ becomes very large, indicating that homodyne measurement approaches full optimality in this limit. To clarify this, in Fig.~\ref{fig:gaussian_thermometry}(c), we depict $m_0$ as a function of $\Delta/T_0$ for both low and high temperature regimes. Increasing width is generally shown to result in $m_0$ becoming larger, i.e., the Gaussian measurements becoming more optimal. The behavior for low and high temperatures is opposite. For low temperatures ($\Delta/T_0$ small), photon detectors ($m_0 {=}1$) surpass Gaussian measurements and as $\Delta/T_0$ increases beyond a threshold, the optimal Gaussian homodyne measurement approaches optimality. For high temperatures, as long as the optimal Gaussian measurement remains a heterodyne, increasing the width $\Delta/T_0$ leads this to approach optimality. However, for very large widths $\Delta/T_0 {{\simeq}} 2$, the optimal Gaussian measurement flips from heterodyne to homodyne,  becoming inferior to photon detectors. \\

\subsection{Gaussian phase estimation} Let us now consider a generic single mode Gaussian state undergoing a phase rotation unitary $U(\lambda){=} \exp(-i\lambda a^{\dagger}a )$, with an unknown phase parameter $\lambda$, with $\lambda{\in}[\lambda_0 {-} \Delta/2, \lambda_0 {+} \Delta/2]$. Since this generic input state can be expressed as a displaced squeezed thermal state, there are five complex parameters completely specifying it, viz., the displacement vector $\vec{d_0} {=} d_0 e^{i\theta}$, the squeezing vector $\vec{s} {=} s e^{i \psi}$, and the thermal photon number $n_{\text{th}}$. Both preparation and measurement parameters are optimized. After undergoing the rotation, the probe state has first moment  $\vec{d} = \sqrt{2}d_0 [\cos \tilde{\theta} ~ \sin \tilde{\theta}]^T$, and second moment
\begin{equation}
 \sigma = \frac{2 n_{\text{th} + 1}}{2} \begin{pmatrix}
       \cosh 2s - \sinh 2s \cos \tilde{\psi}   & -\sinh 2s \sin\tilde{\psi} \\
        -\sinh 2s \sin\tilde{\psi} & \cosh 2s + \sinh 2s \cos\tilde{\psi} 
   \end{pmatrix}, 
   \label{eq:probe_cov_mat}
\end{equation} 
where $\tilde{\theta} {=} \theta {-} \lambda$, and $\tilde{\psi} {=} \psi {-} 2\lambda$. The mean photon number $n$ of the encoded probe state is given by $n { =} d_0^2{ +} (n_\text{th} {+} 1/2) \cosh 2s{-}1/2\,$
The Gaussian POVM is given by $\Pi(\vec{y_m}) {=} D(\vec{y_m})^{\dagger} \Pi_{0}^{m} D(\vec{y_m})$ \cite{giedke2002characterization}, where $\Pi_{0}^{m}$ is the squeezed vacuum (SV) state with squeezing $\vec{s_m}{=} s_m e^{i\psi_m}$ with $s_m{\rightarrow} 0$ ($s_m{\rightarrow}\infty$) being the heterodyne (homodyne) limits. For calculating FI, we need only optimize over $\sigma_m$. By noting that we can re-assign the measurement squeezing phase to the probe preparation step, one need only consider general dyne measurements for finding SGU $\mathcal{G}$. Let us illustrate the results for specific probe preparations. 
\noindent

\subsubsection{Displaced vacuum probes} Let us first consider displaced vacuum probes, i.e., $s {=} n_\text{th} {=} 0$. The CFI is given by $F_{C} {=} 4d_0^2 \left[1{-}\tanh(s_m)\cos \chi \right]$, where $\chi {=}2\theta {-} 2\lambda {-} \psi_m$. The optimal measurement for local sensing is thus a homodyne ($s_m{\rightarrow}\infty$), when $\chi {=} \pi$. SGU $\mathcal{G}$ can then be expressed as $\mathcal{G}(\chi_0, \Delta) {=} \min_{s_m} d_0^2\Delta\cosh{s_m}\sum_{\pm} \arctan\left(e^{s_m} \tan \frac{\Delta{\pm}\chi_0}{2} \right)$. One can show that for finite width $\Delta$, this minimization occurs for $s_m {=} 0$. Thus, in remarkable contrast to the local estimation scenario where homodyne is the optimal strategy \cite{oh2019optimal}, heterodyne is in fact always the optimal measurement strategy in the global setting. Moreover, the SGU $\mathcal{G}(\chi_0, \Delta) {=}1/4d_0^2$, singling out the optimal probe as a highly displaced vacuum state. 

\subsubsection{Squeezed vacuum (SV) probes} 
\begin{figure}
    \centering
    \includegraphics[width = 0.5\textwidth]{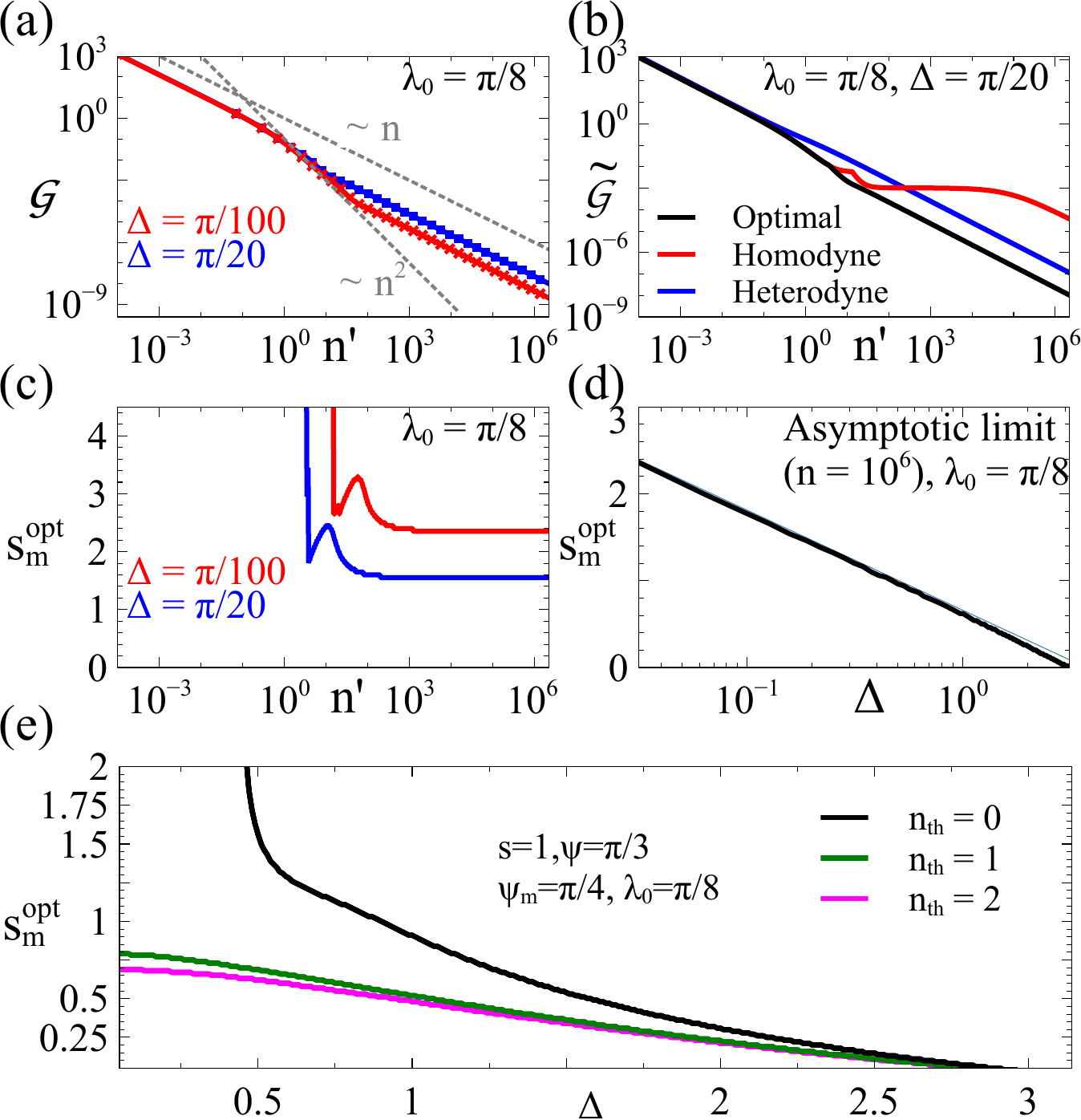}
    \caption{ (a)-(d) for SV probe, (e) for ST probe. (a) Scaling of SGU $\mathcal{G}$ with with $n' {=} n{-}\frac{1}{2}$ for widths $\Delta {=} \pi/100 \text{(red)}, \pi/20 \text{(blue)}$, grey dashed lines show linear and quadratic scalings. (b) scaling of unoptimized $\tilde{\mathcal{G}}$ with $n'$ for $\Delta {=} \pi/20 $ for optimal measurement (black) vs homodyne (red) and heterodyne (blue) measurements, (c) Optimal measurement $s_m^{\text{opt}}$ for (a) with same color scheme. (d) decay of $s_m^{\text{opt}}$ with $\Delta$ in asymptotic limit. (e) $s_m^{\text{opt}}$ with $\Delta$ for ST probes ($s{=}1$) with mean thermal photon number $n_{\text{th}}$. $\lambda_0 {=} \pi/8$ throughout. }
    \label{fig:gaussian_pe}
\end{figure}

Let us now assume that the probe is initially prepared in a SV state, i.e., we set $d_0 {=}  n_{\text{th}} {=} 0$.The expression of CFI for such SV probes with respect to any rotation angle $\lambda$ is given as 
\begin{widetext}
    \begin{equation}
    F_{C} = 4(2n+1)^2 \tanh^2 2s \frac{(4n+2) \cosh 2s_m + \cosh^2 2s_m + (2n+1)^2 \sech^2 2s -\sinh 2s_m (\cos 2\chi \sinh 2s_m + (4n+2) \cos \chi \tanh 2s)}{\left[(4n+2)\cosh 2s_m + \cosh^2 2s_m + (2n+1)^2 \sech^2 2s - \sinh 2s_m (\sinh 2s_m + (4n+2) \cos \chi \tanh 2s)\right]^2}, 
    \label{eq:fc_squeezed vacuum}
\end{equation}
\end{widetext}

\noindent where $\chi {=} 2\lambda {-} \psi{+ }\psi_m$. For local sensing  of the rotation angle $\lambda$, where  $\Delta{\rightarrow}0$, it is well known that the QFI with respect to $\lambda$ scales quadratically with mean photon number $n$ beyond a threshold degree of squeezing \cite{giovannetti2004quantum,oh2019optimal}. Thus it is natural to wonder how such quantum-enhanced sensitivity changes in the domain of global sensing via SGU. In Fig.~\ref{fig:gaussian_pe}(a), we show that even when optimized over the preparation squeezing phases $\psi$, this is generically not the case asymptotically. As the figure shows, for local sensing ($\Delta{\rightarrow}0$), the SGU scales as ${\sim} 1/n^2$ as the mean photon number $n$ goes beyond $n{{\sim}} 1$. For finite $\Delta$, the SGU initially follows quantum-enhanced $ {\sim} 1/n^2$ scaling before reverting to ${{\sim}} 1/n$ standard limit for very strong squeezing $s$, i.e., $n {\gg} 1$. Let us now concentrate on the optimal measurement characterized by $s_m {=} s_m^{\text{opt}}$. In Fig.~\ref{fig:gaussian_pe}(b), we plot $\tilde{\mathcal{G}}$ as a function of mean photon number $n$ for three different measurement setups - viz., optimal Gaussian measurement $s_m^{\text{opt}}$ where $\tilde{\mathcal{G}}{=}\mathcal{G}$, homodyne $s_m {\rightarrow}\infty$, and heterodyne $s_m {=} 0$. As it is evident from the figure, generically neither homodyne nor heterodyne are optimal Gaussian measurements. Moreover, for finite $\Delta$, as $n$ increases, the heterodyne measurement outperforms homodyne, though neither is optimal. This is in stark contrast to the local sensing case, where homodyne is optimal throughout. In Fig.~\ref{fig:gaussian_pe}(c), we plot the optimal measurement parameter $s_m^{\text{opt}}$ as a function of $n$ for $\lambda_0 {=} \pi/8$ and two different choices of width $\Delta$. The figure clearly shows that the optimal measurement starts from homodyne ($s_m^{\text{opt}}{\rightarrow} \infty$) for small $n$ and saturates to a general-dyne measurement with finite $s_m^{\text{opt}}$. In addition, the saturation value of $s_m^{\text{opt}}$ decreases as the width $\Delta$ increases. To see the  asymptotic behavior at large $n$, in Fig.~\ref{fig:gaussian_pe}(d) we plot $s_m^{\text{opt}}$ as a function of $\Delta$ when $n{=}10^6$. The optimal measurement $s_m^{\text{opt}}$ scales as $s_m^{\text{opt}} {\sim} \log (1/\Delta)$, which shows as width $\Delta{\rightarrow}\infty$, the heterodyne is the asymptotically optimal measurement.

\subsubsection{Squeezed thermal (ST) probes} Let us now study the effect of thermal photons on the SV probes, i.e., $n_{\text{th}} \neq 0$. The expression of CFI  for squeezed thermal state probes is given by Eq.~\eqref{eq:fc_squeezed vacuum} again, with the difference that $n$ is now the sum of mean photon number corresponding to probe squeezing $\bar{n}$, and the thermal photon number $n_{\text{thermal}}$. For simplicity, we do not optimize over phase angles $\psi, \psi_m$ in this case. In Fig.~\ref{fig:gaussian_pe}(e), we plot $s_m^{\text{opt}}$ as a function of width $\Delta$ and illustrate that as the number of thermal photons increase, the optimal measurement squeezing $s_m^{\text{opt}}$ for a specified width $\Delta \lambda$ always decreases. Thus, the presence of thermal photons pushes the optimal measurements even more strongly away from the homodyne limit optimal for local estimation, and eventually approaching the heterodyne asymptotically.

\subsection{Multimode generalization} For simplicity, in the above examples, we have only considered single mode probes. However, the analytical expression for CFI in Eq.~\eqref{eq:fc_gaussian} is general to any N-mode Gaussian system, where preparation and measurement covariance matrices $\sigma$ and $\sigma_m$, being $2N{\times} 2N$ real symmetric matrices, have at most $\mathcal{O}(N^2)$ parameters. Thus the optimization for multi-mode SGU is numerically scalable which will be treated elsewhere.
\section{Fermionic Gaussian Global Sensing} To show the generality of our formalism, we now focus on free fermionic quantum probes \cite{mukhopadhyay2024modular,yu2025experimental,mukhopadhyay2025current}. A free fermionic probe of $N$ particles, encoding a parameter $\lambda$, can in general be described by a fermionic Gaussian Hamiltonians $H$ of the form $H(\lambda){=}\Psi^{\dagger} \tilde{H}(\lambda)\Psi$, with $\tilde{H}(\lambda)$ being a $2N{\times}2N$ Hermitian matrix and $\Psi$ being the collection of fermionic creation and annihilation operators. This Hamiltonian can be diagonalized in the momentum space through Bogoliubov transformation  as $H {=}\bigoplus_{k} d_{k}^{\dagger} W_{k}d_{k}$, where $d_k$ are the transformed excitations and $W_k$'s are Hermitian matrices of equal size, whose sizes are given by the periodicity of the lattice in position space, for example $W_k$'s are all $2{\times}2$ matrices for uniform lattices. Thus, the ground state $|\psi_{\text{g}}\rangle$ can be written down in the momentum space in the product form $|\psi_{\text{g}}\rangle {=} \bigotimes_{k} |\psi_k\rangle$, and consequently one adopts the ansatz that the optimal measurement basis $\Pi (\lambda)$ is a product of measurement bases $\lbrace\Pi_k\rbrace$ for each momentum-space cell. Thus, noting that the probe preparation parameters can be rescaled and absorbed into the expression of unknown parameter $\lambda$, the SGU optimization reads $\mathcal{G} {=} \min_{\Pi} \tilde{\mathcal{G}} {=} \min_{\lbrace \Pi_k\rbrace} \int_{\lambda_0{-}\frac{\Delta}{2}}^{\lambda_0{+}\frac{\Delta}{2}}\frac{d\lambda'}{\sum_{k} F_{c} (\lambda'{,} \Pi_k)}$, where we used the additivity of FI for product states. Let us now assume each momentum unit-cell $k$ has measurement parameters $\lbrace \mu^{ik}_{m} \rbrace$. For example, qubit unit cells would each have two measurement basis parameters, the azimuthal and polar angles, i.e., $\mu^{1k}_{m}{=}\theta_m^k, \mu^{2k}_{m}{=}\phi^k_m$. Then, the optimal measurement strategy reduces to solving the system $\partial_{\mu^{ik}_m}\tilde{\mathcal{G}}(\lambda_0, \Delta,\mu^{ik}_m){=}0$ for all discrete momenta $k$ in the first Brillouin zone. Thus, the computational overhead of finding SGU scales at worst linearly, even though the Hilbert space grows exponentially. As we demonstrate below for a canonical example of many-body ground state sensing, this optimization can be even simpler.

\subsection{Transverse XY-chain magnetometry} Let us consider an $XY$ probe in an external transverse magnetic field $h$~\cite{lieb1961two}, with the Hamiltonian $H{=} J \sum_{i} \left[ \left(\frac{1+\gamma}{2}\right) \sigma_i^x \sigma^{x}_{i+1} {+} \left(\frac{1-\gamma}{2}\right) \sigma_i^y \sigma^{y}_{i+1}\right] {+} h \sum_{i} \sigma^z_i$, where ${J}$ is the known nearest-neighbour coupling strength, and $\gamma$ is the anisotropy of the probe. By denoting $\lambda{=}h/J$, the magnetometry problem turns into one of estimating $\lambda$. After a Jordan-Wigner transformation, this Hamiltonian can be written in the momentum-space as $H {=} \bigoplus_k \begin{pmatrix}
    \lambda {-} \cos k & -i\gamma e^{-2i\phi} \sin k \\
    i\gamma e^{2i\phi} \sin k  & {-}\lambda {+} \cos k
\end{pmatrix}$. With gauge choice $\phi{=}\pi/4$, this Hamiltonian can be written down in the new pseudospin $x{-}z$-plane as $H{=}\bigoplus_k \left[(\lambda {-}\cos k) \sigma^z{-}\gamma\sin k\sigma^x \right]$, leading to the (unnormalized) ground state $|\psi_g\rangle{=}\bigotimes_k \left( \cos\theta_k |0\rangle_k + \sin \theta_k|1\rangle_k\right)$, where $\theta_k$ encodes the information about the unknown parameter $\lambda$ as well as anisotropy $\gamma$. One may check that any measurement basis lying on this $x{-}z$ pseudospin basis maximizes the CFI with respect to $\lambda$. That is, the measurement basis for saturating the CR inequality in local sensing paradigm is independent of unknown parameter value $\lambda$. Thus, for ground state sensing with transverse-XY probes, the formulation in Ref.~\cite{montenegro2021global} coincides with our SGU approach, with $\mathcal{G}$ equalling $G$ in Eq.~\eqref{eq:avg_uncertainty}. Of course, for more generic free-fermionic models this may not prevail. However, at least for low number of excitations, we can consider only those momenta cell upto a constant cutoff $k {<} k_0$, and thus compute SGU with constant overhead irrespective of lattice sizes. \\

\section{Conclusion} In this paper, we have introduced a new operationally motivated figure of merit, i.e., SGU, for global sensing, to find a saturable precision bound within the fixed-strategy frequentist paradigm. Our construction also allows simultaneous optimization of probe preparation and measurement strategies. We  demonstrated that the SGU can be calculated efficiently for both bosonic Gaussian and many-body free-fermionic (that is, fermionic Gaussian) platforms, representing two pillars of quantum sensor design. For various bosonic Gaussian probes, we identify a common trend that by increasing the parameter width the optimal measurement transforms from homodyne to heterodyne. In the case of free fermions, the optimal measurement is model dependent. For instance, in the case of transverse XY-probe the optimal measurement is independent of the parameter window. A natural generalization of our formalism is to extend  it to the case of multiparameter global estimation, where the CR inequality is generally not saturable even in the local case \cite{suzuki2016explicit,gessner2018sensitivity,albarelli2019evaluating,razavian2020quantumness,yang2025overcoming}.  Moreover, while our optimization procedure over all measurement bases is quasi-analytical,  AI-based approaches \cite{Benedetti21,belliardo2024applications,xu2025toward} may become useful for more complicated sensor probe arrangements. 

\begin{acknowledgements}
Authors are grateful for feedback from an anonymous referee as well as discussions with George Mihailescu, Jesus Rubio, Victor Montenegro, Stefano Olivares, and acknowledge support from the National Key R\&D Program of China (Grant No.
2018YFA0306703), the National Natural Science Foundation of China (Grants No. 12050410253, No. 92065115, and No. 12274059), the Ministry of Science and Technology of China (Grant No. QNJ2021167001L), and the EU and MIUR through the project PRIN22-2022T25TR3-RISQUE.
\end{acknowledgements}

\bibliography{gaussian_global}

\providecommand{\noopsort}[1]{}\providecommand{\singleletter}[1]{#1}%
\begin{thebibliography}{99}%
\makeatletter
\providecommand \@ifxundefined [1]{%
 \@ifx{#1\undefined}
}%
\providecommand \@ifnum [1]{%
 \ifnum #1\expandafter \@firstoftwo
 \else \expandafter \@secondoftwo
 \fi
}%
\providecommand \@ifx [1]{%
 \ifx #1\expandafter \@firstoftwo
 \else \expandafter \@secondoftwo
 \fi
}%
\providecommand \natexlab [1]{#1}%
\providecommand \enquote  [1]{``#1''}%
\providecommand \bibnamefont  [1]{#1}%
\providecommand \bibfnamefont [1]{#1}%
\providecommand \citenamefont [1]{#1}%
\providecommand \href@noop [0]{\@secondoftwo}%
\providecommand \href [0]{\begingroup \@sanitize@url \@href}%
\providecommand \@href[1]{\@@startlink{#1}\@@href}%
\providecommand \@@href[1]{\endgroup#1\@@endlink}%
\providecommand \@sanitize@url [0]{\catcode `\\12\catcode `\$12\catcode `\&12\catcode `\#12\catcode `\^12\catcode `\_12\catcode `\%12\relax}%
\providecommand \@@startlink[1]{}%
\providecommand \@@endlink[0]{}%
\providecommand \url  [0]{\begingroup\@sanitize@url \@url }%
\providecommand \@url [1]{\endgroup\@href {#1}{\urlprefix }}%
\providecommand \urlprefix  [0]{URL }%
\providecommand \Eprint [0]{\href }%
\providecommand \doibase [0]{https://doi.org/}%
\providecommand \selectlanguage [0]{\@gobble}%
\providecommand \bibinfo  [0]{\@secondoftwo}%
\providecommand \bibfield  [0]{\@secondoftwo}%
\providecommand \translation [1]{[#1]}%
\providecommand \BibitemOpen [0]{}%
\providecommand \bibitemStop [0]{}%
\providecommand \bibitemNoStop [0]{.\EOS\space}%
\providecommand \EOS [0]{\spacefactor3000\relax}%
\providecommand \BibitemShut  [1]{\csname bibitem#1\endcsname}%
\let\auto@bib@innerbib\@empty
\bibitem [{\citenamefont {Giovannetti}\ \emph {et~al.}(2004)\citenamefont {Giovannetti}, \citenamefont {Lloyd},\ and\ \citenamefont {Maccone}}]{giovannetti2004quantum}%
  \BibitemOpen
  \bibfield  {author} {\bibinfo {author} {\bibfnamefont {V.}~\bibnamefont {Giovannetti}}, \bibinfo {author} {\bibfnamefont {S.}~\bibnamefont {Lloyd}},\ and\ \bibinfo {author} {\bibfnamefont {L.}~\bibnamefont {Maccone}},\ }\bibfield  {title} {\bibinfo {title} {Quantum-enhanced measurements: beating the standard quantum limit},\ }\href@noop {} {\bibfield  {journal} {\bibinfo  {journal} {Science}\ }\textbf {\bibinfo {volume} {306}},\ \bibinfo {pages} {1330} (\bibinfo {year} {2004})}\BibitemShut {NoStop}%
\bibitem [{\citenamefont {Giovannetti}\ \emph {et~al.}(2006)\citenamefont {Giovannetti}, \citenamefont {Lloyd},\ and\ \citenamefont {Maccone}}]{giovannetti2006quantum}%
  \BibitemOpen
  \bibfield  {author} {\bibinfo {author} {\bibfnamefont {V.}~\bibnamefont {Giovannetti}}, \bibinfo {author} {\bibfnamefont {S.}~\bibnamefont {Lloyd}},\ and\ \bibinfo {author} {\bibfnamefont {L.}~\bibnamefont {Maccone}},\ }\bibfield  {title} {\bibinfo {title} {Quantum metrology},\ }\href@noop {} {\bibfield  {journal} {\bibinfo  {journal} {Physical review letters}\ }\textbf {\bibinfo {volume} {96}},\ \bibinfo {pages} {010401} (\bibinfo {year} {2006})}\BibitemShut {NoStop}%
\bibitem [{\citenamefont {Giovannetti}\ \emph {et~al.}(2011)\citenamefont {Giovannetti}, \citenamefont {Lloyd},\ and\ \citenamefont {Maccone}}]{giovannetti2011advances}%
  \BibitemOpen
  \bibfield  {author} {\bibinfo {author} {\bibfnamefont {V.}~\bibnamefont {Giovannetti}}, \bibinfo {author} {\bibfnamefont {S.}~\bibnamefont {Lloyd}},\ and\ \bibinfo {author} {\bibfnamefont {L.}~\bibnamefont {Maccone}},\ }\bibfield  {title} {\bibinfo {title} {Advances in quantum metrology},\ }\href@noop {} {\bibfield  {journal} {\bibinfo  {journal} {Nature photonics}\ }\textbf {\bibinfo {volume} {5}},\ \bibinfo {pages} {222} (\bibinfo {year} {2011})}\BibitemShut {NoStop}%
\bibitem [{\citenamefont {T{\'o}th}\ and\ \citenamefont {Apellaniz}(2014)}]{toth2014quantum}%
  \BibitemOpen
  \bibfield  {author} {\bibinfo {author} {\bibfnamefont {G.}~\bibnamefont {T{\'o}th}}\ and\ \bibinfo {author} {\bibfnamefont {I.}~\bibnamefont {Apellaniz}},\ }\bibfield  {title} {\bibinfo {title} {Quantum metrology from a quantum information science perspective},\ }\href@noop {} {\bibfield  {journal} {\bibinfo  {journal} {Journal of Physics A: Mathematical and Theoretical}\ }\textbf {\bibinfo {volume} {47}},\ \bibinfo {pages} {424006} (\bibinfo {year} {2014})}\BibitemShut {NoStop}%
\bibitem [{\citenamefont {Paris}(2015)}]{Paris_2016}%
  \BibitemOpen
  \bibfield  {author} {\bibinfo {author} {\bibfnamefont {M.~G.~A.}\ \bibnamefont {Paris}},\ }\bibfield  {title} {\bibinfo {title} {Achieving the landau bound to precision of quantum thermometry in systems with vanishing gap},\ }\href@noop {} {\bibfield  {journal} {\bibinfo  {journal} {Journal of Physics A: Mathematical and Theoretical}\ }\textbf {\bibinfo {volume} {49}},\ \bibinfo {pages} {03LT02} (\bibinfo {year} {2015})}\BibitemShut {NoStop}%
\bibitem [{\citenamefont {Qvarfort}\ \emph {et~al.}(2018)\citenamefont {Qvarfort}, \citenamefont {Serafini}, \citenamefont {Barker},\ and\ \citenamefont {Bose}}]{qvarfort2018gravimetry}%
  \BibitemOpen
  \bibfield  {author} {\bibinfo {author} {\bibfnamefont {S.}~\bibnamefont {Qvarfort}}, \bibinfo {author} {\bibfnamefont {A.}~\bibnamefont {Serafini}}, \bibinfo {author} {\bibfnamefont {P.~F.}\ \bibnamefont {Barker}},\ and\ \bibinfo {author} {\bibfnamefont {S.}~\bibnamefont {Bose}},\ }\bibfield  {title} {\bibinfo {title} {Gravimetry through non-linear optomechanics},\ }\href@noop {} {\bibfield  {journal} {\bibinfo  {journal} {Nature communications}\ }\textbf {\bibinfo {volume} {9}},\ \bibinfo {pages} {3690} (\bibinfo {year} {2018})}\BibitemShut {NoStop}%
\bibitem [{\citenamefont {Montenegro}\ \emph {et~al.}(2024)\citenamefont {Montenegro}, \citenamefont {Mukhopadhyay}, \citenamefont {Yousefjani}, \citenamefont {Sarkar}, \citenamefont {Mishra}, \citenamefont {Paris},\ and\ \citenamefont {Bayat}}]{montenegro2024quantum}%
  \BibitemOpen
  \bibfield  {author} {\bibinfo {author} {\bibfnamefont {V.}~\bibnamefont {Montenegro}}, \bibinfo {author} {\bibfnamefont {C.}~\bibnamefont {Mukhopadhyay}}, \bibinfo {author} {\bibfnamefont {R.}~\bibnamefont {Yousefjani}}, \bibinfo {author} {\bibfnamefont {S.}~\bibnamefont {Sarkar}}, \bibinfo {author} {\bibfnamefont {U.}~\bibnamefont {Mishra}}, \bibinfo {author} {\bibfnamefont {M.~G.~A.}\ \bibnamefont {Paris}},\ and\ \bibinfo {author} {\bibfnamefont {A.}~\bibnamefont {Bayat}},\ }\bibfield  {title} {\bibinfo {title} {Quantum metrology and sensing with many-body systems},\ }\href@noop {} {\bibfield  {journal} {\bibinfo  {journal} {arXiv preprint arXiv:2408.15323}\ } (\bibinfo {year} {2024})}\BibitemShut {NoStop}%
\bibitem [{\citenamefont {He}\ \emph {et~al.}(2023)\citenamefont {He}, \citenamefont {Yousefjani},\ and\ \citenamefont {Bayat}}]{he2023stark}%
  \BibitemOpen
  \bibfield  {author} {\bibinfo {author} {\bibfnamefont {X.}~\bibnamefont {He}}, \bibinfo {author} {\bibfnamefont {R.}~\bibnamefont {Yousefjani}},\ and\ \bibinfo {author} {\bibfnamefont {A.}~\bibnamefont {Bayat}},\ }\bibfield  {title} {\bibinfo {title} {Stark localization as a resource for weak-field sensing with super-heisenberg precision},\ }\href@noop {} {\bibfield  {journal} {\bibinfo  {journal} {Phys. Rev. Lett.}\ }\textbf {\bibinfo {volume} {131}},\ \bibinfo {pages} {010801} (\bibinfo {year} {2023})}\BibitemShut {NoStop}%
\bibitem [{\citenamefont {Sarkar}\ \emph {et~al.}(2022)\citenamefont {Sarkar}, \citenamefont {Mukhopadhyay}, \citenamefont {Alase},\ and\ \citenamefont {Bayat}}]{free2022sarkar}%
  \BibitemOpen
  \bibfield  {author} {\bibinfo {author} {\bibfnamefont {S.}~\bibnamefont {Sarkar}}, \bibinfo {author} {\bibfnamefont {C.}~\bibnamefont {Mukhopadhyay}}, \bibinfo {author} {\bibfnamefont {A.}~\bibnamefont {Alase}},\ and\ \bibinfo {author} {\bibfnamefont {A.}~\bibnamefont {Bayat}},\ }\bibfield  {title} {\bibinfo {title} {Free-fermionic topological quantum sensors},\ }\href@noop {} {\bibfield  {journal} {\bibinfo  {journal} {Phys. Rev. Lett.}\ }\textbf {\bibinfo {volume} {129}},\ \bibinfo {pages} {090503} (\bibinfo {year} {2022})}\BibitemShut {NoStop}%
\bibitem [{\citenamefont {De~Chiara}\ and\ \citenamefont {Sanpera}(2018)}]{de2018genuine}%
  \BibitemOpen
  \bibfield  {author} {\bibinfo {author} {\bibfnamefont {G.}~\bibnamefont {De~Chiara}}\ and\ \bibinfo {author} {\bibfnamefont {A.}~\bibnamefont {Sanpera}},\ }\bibfield  {title} {\bibinfo {title} {Genuine quantum correlations in quantum many-body systems: a review of recent progress},\ }\href@noop {} {\bibfield  {journal} {\bibinfo  {journal} {Reports on Progress in Physics}\ }\textbf {\bibinfo {volume} {81}},\ \bibinfo {pages} {074002} (\bibinfo {year} {2018})}\BibitemShut {NoStop}%
\bibitem [{\citenamefont {De~Pasquale}\ and\ \citenamefont {Stace}(2018)}]{de2018quantum}%
  \BibitemOpen
  \bibfield  {author} {\bibinfo {author} {\bibfnamefont {A.}~\bibnamefont {De~Pasquale}}\ and\ \bibinfo {author} {\bibfnamefont {T.~M.}\ \bibnamefont {Stace}},\ }\bibfield  {title} {\bibinfo {title} {Quantum thermometry},\ }in\ \href@noop {} {\emph {\bibinfo {booktitle} {Thermodynamics in the quantum regime}}}\ (\bibinfo  {publisher} {Springer},\ \bibinfo {year} {2018})\ pp.\ \bibinfo {pages} {503--527}\BibitemShut {NoStop}%
\bibitem [{\citenamefont {Correa}\ \emph {et~al.}(2015)\citenamefont {Correa}, \citenamefont {Mehboudi}, \citenamefont {Adesso},\ and\ \citenamefont {Sanpera}}]{correa2015individual}%
  \BibitemOpen
  \bibfield  {author} {\bibinfo {author} {\bibfnamefont {L.~A.}\ \bibnamefont {Correa}}, \bibinfo {author} {\bibfnamefont {M.}~\bibnamefont {Mehboudi}}, \bibinfo {author} {\bibfnamefont {G.}~\bibnamefont {Adesso}},\ and\ \bibinfo {author} {\bibfnamefont {A.}~\bibnamefont {Sanpera}},\ }\bibfield  {title} {\bibinfo {title} {Individual quantum probes for optimal thermometry},\ }\href@noop {} {\bibfield  {journal} {\bibinfo  {journal} {Physical review letters}\ }\textbf {\bibinfo {volume} {114}},\ \bibinfo {pages} {220405} (\bibinfo {year} {2015})}\BibitemShut {NoStop}%
\bibitem [{\citenamefont {Mehboudi}\ \emph {et~al.}(2019)\citenamefont {Mehboudi}, \citenamefont {Sanpera},\ and\ \citenamefont {Correa}}]{mehboudi2019thermometry}%
  \BibitemOpen
  \bibfield  {author} {\bibinfo {author} {\bibfnamefont {M.}~\bibnamefont {Mehboudi}}, \bibinfo {author} {\bibfnamefont {A.}~\bibnamefont {Sanpera}},\ and\ \bibinfo {author} {\bibfnamefont {L.~A.}\ \bibnamefont {Correa}},\ }\bibfield  {title} {\bibinfo {title} {Thermometry in the quantum regime: recent theoretical progress},\ }\href@noop {} {\bibfield  {journal} {\bibinfo  {journal} {Journal of Physics A: Mathematical and Theoretical}\ }\textbf {\bibinfo {volume} {52}},\ \bibinfo {pages} {303001} (\bibinfo {year} {2019})}\BibitemShut {NoStop}%
\bibitem [{\citenamefont {Mukhopadhyay}\ \emph {et~al.}(2018)\citenamefont {Mukhopadhyay}, \citenamefont {Gupta},\ and\ \citenamefont {Pati}}]{mukhopadhyay2018superposition}%
  \BibitemOpen
  \bibfield  {author} {\bibinfo {author} {\bibfnamefont {C.}~\bibnamefont {Mukhopadhyay}}, \bibinfo {author} {\bibfnamefont {M.~K.}\ \bibnamefont {Gupta}},\ and\ \bibinfo {author} {\bibfnamefont {A.~K.}\ \bibnamefont {Pati}},\ }\bibfield  {title} {\bibinfo {title} {Superposition of causal order as a metrological resource for quantum thermometry},\ }\href@noop {} {\bibfield  {journal} {\bibinfo  {journal} {arXiv preprint arXiv:1812.07508}\ } (\bibinfo {year} {2018})}\BibitemShut {NoStop}%
\bibitem [{\citenamefont {Pati}\ \emph {et~al.}(2020)\citenamefont {Pati}, \citenamefont {Mukhopadhyay}, \citenamefont {Chakraborty},\ and\ \citenamefont {Ghosh}}]{pati2020quantum}%
  \BibitemOpen
  \bibfield  {author} {\bibinfo {author} {\bibfnamefont {A.~K.}\ \bibnamefont {Pati}}, \bibinfo {author} {\bibfnamefont {C.}~\bibnamefont {Mukhopadhyay}}, \bibinfo {author} {\bibfnamefont {S.}~\bibnamefont {Chakraborty}},\ and\ \bibinfo {author} {\bibfnamefont {S.}~\bibnamefont {Ghosh}},\ }\bibfield  {title} {\bibinfo {title} {Quantum precision thermometry with weak measurements},\ }\href@noop {} {\bibfield  {journal} {\bibinfo  {journal} {Physical Review A}\ }\textbf {\bibinfo {volume} {102}},\ \bibinfo {pages} {012204} (\bibinfo {year} {2020})}\BibitemShut {NoStop}%
\bibitem [{\citenamefont {Zhao}\ \emph {et~al.}(2020)\citenamefont {Zhao}, \citenamefont {Yang},\ and\ \citenamefont {Chiribella}}]{zhao2020quantum}%
  \BibitemOpen
  \bibfield  {author} {\bibinfo {author} {\bibfnamefont {X.}~\bibnamefont {Zhao}}, \bibinfo {author} {\bibfnamefont {Y.}~\bibnamefont {Yang}},\ and\ \bibinfo {author} {\bibfnamefont {G.}~\bibnamefont {Chiribella}},\ }\bibfield  {title} {\bibinfo {title} {Quantum metrology with indefinite causal order},\ }\href@noop {} {\bibfield  {journal} {\bibinfo  {journal} {Physical Review Letters}\ }\textbf {\bibinfo {volume} {124}},\ \bibinfo {pages} {190503} (\bibinfo {year} {2020})}\BibitemShut {NoStop}%
\bibitem [{\citenamefont {Aslam}\ \emph {et~al.}(2023)\citenamefont {Aslam}, \citenamefont {Zhou}, \citenamefont {Urbach}, \citenamefont {Turner}, \citenamefont {Walsworth}, \citenamefont {Lukin},\ and\ \citenamefont {Park}}]{aslam2023quantum}%
  \BibitemOpen
  \bibfield  {author} {\bibinfo {author} {\bibfnamefont {N.}~\bibnamefont {Aslam}}, \bibinfo {author} {\bibfnamefont {H.}~\bibnamefont {Zhou}}, \bibinfo {author} {\bibfnamefont {E.~K.}\ \bibnamefont {Urbach}}, \bibinfo {author} {\bibfnamefont {M.~J.}\ \bibnamefont {Turner}}, \bibinfo {author} {\bibfnamefont {R.~L.}\ \bibnamefont {Walsworth}}, \bibinfo {author} {\bibfnamefont {M.~D.}\ \bibnamefont {Lukin}},\ and\ \bibinfo {author} {\bibfnamefont {H.}~\bibnamefont {Park}},\ }\bibfield  {title} {\bibinfo {title} {Quantum sensors for biomedical applications},\ }\href@noop {} {\bibfield  {journal} {\bibinfo  {journal} {Nature Reviews Physics}\ }\textbf {\bibinfo {volume} {5}},\ \bibinfo {pages} {157} (\bibinfo {year} {2023})}\BibitemShut {NoStop}%
\bibitem [{\citenamefont {Ye}\ and\ \citenamefont {Zoller}(2024)}]{ye2024essay}%
  \BibitemOpen
  \bibfield  {author} {\bibinfo {author} {\bibfnamefont {J.}~\bibnamefont {Ye}}\ and\ \bibinfo {author} {\bibfnamefont {P.}~\bibnamefont {Zoller}},\ }\bibfield  {title} {\bibinfo {title} {Essay: Quantum sensing with atomic, molecular, and optical platforms for fundamental physics},\ }\href@noop {} {\bibfield  {journal} {\bibinfo  {journal} {Physical Review Letters}\ }\textbf {\bibinfo {volume} {132}},\ \bibinfo {pages} {190001} (\bibinfo {year} {2024})}\BibitemShut {NoStop}%
\bibitem [{\citenamefont {Yu}\ \emph {et~al.}(2024)\citenamefont {Yu}, \citenamefont {Li}, \citenamefont {Chu}, \citenamefont {Mera}, \citenamefont {{\"U}nal}, \citenamefont {Yang}, \citenamefont {Liu}, \citenamefont {Goldman},\ and\ \citenamefont {Cai}}]{yu2024experimental}%
  \BibitemOpen
  \bibfield  {author} {\bibinfo {author} {\bibfnamefont {M.}~\bibnamefont {Yu}}, \bibinfo {author} {\bibfnamefont {X.}~\bibnamefont {Li}}, \bibinfo {author} {\bibfnamefont {Y.}~\bibnamefont {Chu}}, \bibinfo {author} {\bibfnamefont {B.}~\bibnamefont {Mera}}, \bibinfo {author} {\bibfnamefont {F.~N.}\ \bibnamefont {{\"U}nal}}, \bibinfo {author} {\bibfnamefont {P.}~\bibnamefont {Yang}}, \bibinfo {author} {\bibfnamefont {Y.}~\bibnamefont {Liu}}, \bibinfo {author} {\bibfnamefont {N.}~\bibnamefont {Goldman}},\ and\ \bibinfo {author} {\bibfnamefont {J.}~\bibnamefont {Cai}},\ }\bibfield  {title} {\bibinfo {title} {Experimental demonstration of topological bounds in quantum metrology},\ }\href@noop {} {\bibfield  {journal} {\bibinfo  {journal} {National Science Review}\ ,\ \bibinfo {pages} {nwae065}} (\bibinfo {year} {2024})}\BibitemShut {NoStop}%
\bibitem [{\citenamefont {Gribben}\ \emph {et~al.}(2024)\citenamefont {Gribben}, \citenamefont {Sanpera}, \citenamefont {Fazio}, \citenamefont {Marino},\ and\ \citenamefont {Iemini}}]{gribben2024quantum}%
  \BibitemOpen
  \bibfield  {author} {\bibinfo {author} {\bibfnamefont {D.}~\bibnamefont {Gribben}}, \bibinfo {author} {\bibfnamefont {A.}~\bibnamefont {Sanpera}}, \bibinfo {author} {\bibfnamefont {R.}~\bibnamefont {Fazio}}, \bibinfo {author} {\bibfnamefont {J.}~\bibnamefont {Marino}},\ and\ \bibinfo {author} {\bibfnamefont {F.}~\bibnamefont {Iemini}},\ }\bibfield  {title} {\bibinfo {title} {Quantum enhancements and entropic constraints to boundary time crystals as sensors of ac fields},\ }\href@noop {} {\bibfield  {journal} {\bibinfo  {journal} {arXiv preprint arXiv:2406.06273}\ } (\bibinfo {year} {2024})}\BibitemShut {NoStop}%
\bibitem [{\citenamefont {Cram{\'e}r}(1999)}]{cramer1999mathematical}%
  \BibitemOpen
  \bibfield  {author} {\bibinfo {author} {\bibfnamefont {H.}~\bibnamefont {Cram{\'e}r}},\ }\href@noop {} {\emph {\bibinfo {title} {Mathematical methods of statistics}}},\ Vol.~\bibinfo {volume} {26}\ (\bibinfo  {publisher} {Princeton university press},\ \bibinfo {year} {1999})\BibitemShut {NoStop}%
\bibitem [{\citenamefont {Rao}(1992)}]{rao1992information}%
  \BibitemOpen
  \bibfield  {author} {\bibinfo {author} {\bibfnamefont {C.~R.}\ \bibnamefont {Rao}},\ }\bibfield  {title} {\bibinfo {title} {Information and the accuracy attainable in the estimation of statistical parameters},\ }in\ \href@noop {} {\emph {\bibinfo {booktitle} {Breakthroughs in Statistics: Foundations and basic theory}}}\ (\bibinfo  {publisher} {Springer},\ \bibinfo {year} {1992})\ pp.\ \bibinfo {pages} {235--247}\BibitemShut {NoStop}%
\bibitem [{\citenamefont {Helstrom}(1969)}]{helstrom1969quantum}%
  \BibitemOpen
  \bibfield  {author} {\bibinfo {author} {\bibfnamefont {C.~W.}\ \bibnamefont {Helstrom}},\ }\bibfield  {title} {\bibinfo {title} {Quantum detection and estimation theory},\ }\href@noop {} {\bibfield  {journal} {\bibinfo  {journal} {Journal of Statistical Physics}\ }\textbf {\bibinfo {volume} {1}},\ \bibinfo {pages} {231} (\bibinfo {year} {1969})}\BibitemShut {NoStop}%
\bibitem [{\citenamefont {Braunstein}\ and\ \citenamefont {Caves}(1994)}]{statistical1994braunstein}%
  \BibitemOpen
  \bibfield  {author} {\bibinfo {author} {\bibfnamefont {S.~L.}\ \bibnamefont {Braunstein}}\ and\ \bibinfo {author} {\bibfnamefont {C.~M.}\ \bibnamefont {Caves}},\ }\bibfield  {title} {\bibinfo {title} {Statistical distance and the geometry of quantum states},\ }\href@noop {} {\bibfield  {journal} {\bibinfo  {journal} {Phys. Rev. Lett.}\ }\textbf {\bibinfo {volume} {72}},\ \bibinfo {pages} {3439} (\bibinfo {year} {1994})}\BibitemShut {NoStop}%
\bibitem [{\citenamefont {Paris}(2009)}]{paris2009quantum}%
  \BibitemOpen
  \bibfield  {author} {\bibinfo {author} {\bibfnamefont {M.~G.~A.}\ \bibnamefont {Paris}},\ }\bibfield  {title} {\bibinfo {title} {Quantum estimation for quantum technology},\ }\href@noop {} {\bibfield  {journal} {\bibinfo  {journal} {International Journal of Quantum Information}\ }\textbf {\bibinfo {volume} {7}},\ \bibinfo {pages} {125} (\bibinfo {year} {2009})}\BibitemShut {NoStop}%
\bibitem [{\citenamefont {Braun}\ \emph {et~al.}(2018)\citenamefont {Braun}, \citenamefont {Adesso}, \citenamefont {Benatti}, \citenamefont {Floreanini}, \citenamefont {Marzolino}, \citenamefont {Mitchell},\ and\ \citenamefont {Pirandola}}]{braun2018quantum}%
  \BibitemOpen
  \bibfield  {author} {\bibinfo {author} {\bibfnamefont {D.}~\bibnamefont {Braun}}, \bibinfo {author} {\bibfnamefont {G.}~\bibnamefont {Adesso}}, \bibinfo {author} {\bibfnamefont {F.}~\bibnamefont {Benatti}}, \bibinfo {author} {\bibfnamefont {R.}~\bibnamefont {Floreanini}}, \bibinfo {author} {\bibfnamefont {U.}~\bibnamefont {Marzolino}}, \bibinfo {author} {\bibfnamefont {M.~W.}\ \bibnamefont {Mitchell}},\ and\ \bibinfo {author} {\bibfnamefont {S.}~\bibnamefont {Pirandola}},\ }\bibfield  {title} {\bibinfo {title} {Quantum-enhanced measurements without entanglement},\ }\href@noop {} {\bibfield  {journal} {\bibinfo  {journal} {Reviews of Modern Physics}\ }\textbf {\bibinfo {volume} {90}},\ \bibinfo {pages} {035006} (\bibinfo {year} {2018})}\BibitemShut {NoStop}%
\bibitem [{\citenamefont {Degen}\ \emph {et~al.}(2017)\citenamefont {Degen}, \citenamefont {Reinhard},\ and\ \citenamefont {Cappellaro}}]{degen2017quantum}%
  \BibitemOpen
  \bibfield  {author} {\bibinfo {author} {\bibfnamefont {C.~L.}\ \bibnamefont {Degen}}, \bibinfo {author} {\bibfnamefont {F.}~\bibnamefont {Reinhard}},\ and\ \bibinfo {author} {\bibfnamefont {P.}~\bibnamefont {Cappellaro}},\ }\bibfield  {title} {\bibinfo {title} {Quantum sensing},\ }\href@noop {} {\bibfield  {journal} {\bibinfo  {journal} {Rev. Mod. Phys.}\ }\textbf {\bibinfo {volume} {89}},\ \bibinfo {pages} {035002} (\bibinfo {year} {2017})}\BibitemShut {NoStop}%
\bibitem [{\citenamefont {Liu}\ \emph {et~al.}(2019)\citenamefont {Liu}, \citenamefont {Yuan}, \citenamefont {Lu},\ and\ \citenamefont {Wang}}]{liu2019quantum}%
  \BibitemOpen
  \bibfield  {author} {\bibinfo {author} {\bibfnamefont {J.}~\bibnamefont {Liu}}, \bibinfo {author} {\bibfnamefont {H.}~\bibnamefont {Yuan}}, \bibinfo {author} {\bibfnamefont {X.-M.}\ \bibnamefont {Lu}},\ and\ \bibinfo {author} {\bibfnamefont {X.}~\bibnamefont {Wang}},\ }\bibfield  {title} {\bibinfo {title} {Quantum fisher information matrix and multiparameter estimation},\ }\href@noop {} {\bibfield  {journal} {\bibinfo  {journal} {Journal of Physics A: Mathematical and Theoretical}\ }\textbf {\bibinfo {volume} {53}},\ \bibinfo {pages} {023001} (\bibinfo {year} {2019})}\BibitemShut {NoStop}%
\bibitem [{\citenamefont {Mukhopadhyay}\ and\ \citenamefont {Bayat}(2024)}]{mukhopadhyay2024modular}%
  \BibitemOpen
  \bibfield  {author} {\bibinfo {author} {\bibfnamefont {C.}~\bibnamefont {Mukhopadhyay}}\ and\ \bibinfo {author} {\bibfnamefont {A.}~\bibnamefont {Bayat}},\ }\bibfield  {title} {\bibinfo {title} {Modular many-body quantum sensors},\ }\href@noop {} {\bibfield  {journal} {\bibinfo  {journal} {Physical Review Letters}\ }\textbf {\bibinfo {volume} {133}},\ \bibinfo {pages} {120601} (\bibinfo {year} {2024})}\BibitemShut {NoStop}%
\bibitem [{\citenamefont {Rubio}\ and\ \citenamefont {Dunningham}(2019)}]{rubio2019quantum}%
  \BibitemOpen
  \bibfield  {author} {\bibinfo {author} {\bibfnamefont {J.}~\bibnamefont {Rubio}}\ and\ \bibinfo {author} {\bibfnamefont {J.}~\bibnamefont {Dunningham}},\ }\bibfield  {title} {\bibinfo {title} {Quantum metrology in the presence of limited data},\ }\href@noop {} {\bibfield  {journal} {\bibinfo  {journal} {New Journal of Physics}\ }\textbf {\bibinfo {volume} {21}},\ \bibinfo {pages} {043037} (\bibinfo {year} {2019})}\BibitemShut {NoStop}%
\bibitem [{\citenamefont {Rubio}\ \emph {et~al.}(2021)\citenamefont {Rubio}, \citenamefont {Anders},\ and\ \citenamefont {Correa}}]{rubio2021global}%
  \BibitemOpen
  \bibfield  {author} {\bibinfo {author} {\bibfnamefont {J.}~\bibnamefont {Rubio}}, \bibinfo {author} {\bibfnamefont {J.}~\bibnamefont {Anders}},\ and\ \bibinfo {author} {\bibfnamefont {L.~A.}\ \bibnamefont {Correa}},\ }\bibfield  {title} {\bibinfo {title} {Global quantum thermometry},\ }\href@noop {} {\bibfield  {journal} {\bibinfo  {journal} {Physical Review Letters}\ }\textbf {\bibinfo {volume} {127}},\ \bibinfo {pages} {190402} (\bibinfo {year} {2021})}\BibitemShut {NoStop}%
\bibitem [{\citenamefont {Rubio}(2022)}]{rubio2022quantum}%
  \BibitemOpen
  \bibfield  {author} {\bibinfo {author} {\bibfnamefont {J.}~\bibnamefont {Rubio}},\ }\bibfield  {title} {\bibinfo {title} {Quantum scale estimation},\ }\href@noop {} {\bibfield  {journal} {\bibinfo  {journal} {Quantum Science and Technology}\ }\textbf {\bibinfo {volume} {8}},\ \bibinfo {pages} {015009} (\bibinfo {year} {2022})}\BibitemShut {NoStop}%
\bibitem [{\citenamefont {Mok}\ \emph {et~al.}(2021)\citenamefont {Mok}, \citenamefont {Bharti}, \citenamefont {Kwek},\ and\ \citenamefont {Bayat}}]{mok2021optimal}%
  \BibitemOpen
  \bibfield  {author} {\bibinfo {author} {\bibfnamefont {W.-K.}\ \bibnamefont {Mok}}, \bibinfo {author} {\bibfnamefont {K.}~\bibnamefont {Bharti}}, \bibinfo {author} {\bibfnamefont {L.-C.}\ \bibnamefont {Kwek}},\ and\ \bibinfo {author} {\bibfnamefont {A.}~\bibnamefont {Bayat}},\ }\bibfield  {title} {\bibinfo {title} {Optimal probes for global quantum thermometry},\ }\href@noop {} {\bibfield  {journal} {\bibinfo  {journal} {Communications physics}\ }\textbf {\bibinfo {volume} {4}},\ \bibinfo {pages} {62} (\bibinfo {year} {2021})}\BibitemShut {NoStop}%
\bibitem [{\citenamefont {Zhou}\ \emph {et~al.}(2024)\citenamefont {Zhou}, \citenamefont {Qiu},\ and\ \citenamefont {Zhang}}]{zhou2024strict}%
  \BibitemOpen
  \bibfield  {author} {\bibinfo {author} {\bibfnamefont {Z.-Y.}\ \bibnamefont {Zhou}}, \bibinfo {author} {\bibfnamefont {J.-T.}\ \bibnamefont {Qiu}},\ and\ \bibinfo {author} {\bibfnamefont {D.-J.}\ \bibnamefont {Zhang}},\ }\bibfield  {title} {\bibinfo {title} {Strict hierarchy of optimal strategies for global estimations: Linking global estimations with local ones},\ }\href@noop {} {\bibfield  {journal} {\bibinfo  {journal} {Physical Review Research}\ }\textbf {\bibinfo {volume} {6}},\ \bibinfo {pages} {L032048} (\bibinfo {year} {2024})}\BibitemShut {NoStop}%
\bibitem [{\citenamefont {Mihailescu}\ \emph {et~al.}(2024)\citenamefont {Mihailescu}, \citenamefont {Campbell},\ and\ \citenamefont {Gietka}}]{mihailescu2024uncertain}%
  \BibitemOpen
  \bibfield  {author} {\bibinfo {author} {\bibfnamefont {G.}~\bibnamefont {Mihailescu}}, \bibinfo {author} {\bibfnamefont {S.}~\bibnamefont {Campbell}},\ and\ \bibinfo {author} {\bibfnamefont {K.}~\bibnamefont {Gietka}},\ }\bibfield  {title} {\bibinfo {title} {Uncertain quantum critical metrology: From single to multi parameter sensing},\ }\href@noop {} {\bibfield  {journal} {\bibinfo  {journal} {arXiv preprint arXiv:2407.19917}\ } (\bibinfo {year} {2024})}\BibitemShut {NoStop}%
\bibitem [{\citenamefont {D'Ariano}\ \emph {et~al.}(1998)\citenamefont {D'Ariano}, \citenamefont {Macchiavello},\ and\ \citenamefont {Sacchi}}]{d1998general}%
  \BibitemOpen
  \bibfield  {author} {\bibinfo {author} {\bibfnamefont {G.}~\bibnamefont {D'Ariano}}, \bibinfo {author} {\bibfnamefont {C.}~\bibnamefont {Macchiavello}},\ and\ \bibinfo {author} {\bibfnamefont {M.}~\bibnamefont {Sacchi}},\ }\bibfield  {title} {\bibinfo {title} {On the general problem of quantum phase estimation},\ }\href@noop {} {\bibfield  {journal} {\bibinfo  {journal} {Physics Letters A}\ }\textbf {\bibinfo {volume} {248}},\ \bibinfo {pages} {103} (\bibinfo {year} {1998})}\BibitemShut {NoStop}%
\bibitem [{\citenamefont {van Trees}(1968)}]{vantrees1968}%
  \BibitemOpen
  \bibfield  {author} {\bibinfo {author} {\bibfnamefont {H.~L.}\ \bibnamefont {van Trees}},\ }\href@noop {} {\emph {\bibinfo {title} {Detection, Estimation and Modulation Theory, Part 1}}}\ (\bibinfo  {publisher} {Wiley \& Sons},\ \bibinfo {year} {1968})\BibitemShut {NoStop}%
\bibitem [{\citenamefont {Jarzyna}\ and\ \citenamefont {Demkowicz-Dobrza{\'n}ski}(2015)}]{jarzyna2015true}%
  \BibitemOpen
  \bibfield  {author} {\bibinfo {author} {\bibfnamefont {M.}~\bibnamefont {Jarzyna}}\ and\ \bibinfo {author} {\bibfnamefont {R.}~\bibnamefont {Demkowicz-Dobrza{\'n}ski}},\ }\bibfield  {title} {\bibinfo {title} {True precision limits in quantum metrology},\ }\href@noop {} {\bibfield  {journal} {\bibinfo  {journal} {New Journal of Physics}\ }\textbf {\bibinfo {volume} {17}},\ \bibinfo {pages} {013010} (\bibinfo {year} {2015})}\BibitemShut {NoStop}%
\bibitem [{\citenamefont {Rubio}\ and\ \citenamefont {Dunningham}(2020)}]{rubio2020bayesian}%
  \BibitemOpen
  \bibfield  {author} {\bibinfo {author} {\bibfnamefont {J.}~\bibnamefont {Rubio}}\ and\ \bibinfo {author} {\bibfnamefont {J.}~\bibnamefont {Dunningham}},\ }\bibfield  {title} {\bibinfo {title} {Bayesian multiparameter quantum metrology with limited data},\ }\href@noop {} {\bibfield  {journal} {\bibinfo  {journal} {Physical Review A}\ }\textbf {\bibinfo {volume} {101}},\ \bibinfo {pages} {032114} (\bibinfo {year} {2020})}\BibitemShut {NoStop}%
\bibitem [{\citenamefont {Cimini}\ \emph {et~al.}(2024)\citenamefont {Cimini}, \citenamefont {Polino}, \citenamefont {Valeri}, \citenamefont {Spagnolo},\ and\ \citenamefont {Sciarrino}}]{cimini2024benchmarking}%
  \BibitemOpen
  \bibfield  {author} {\bibinfo {author} {\bibfnamefont {V.}~\bibnamefont {Cimini}}, \bibinfo {author} {\bibfnamefont {E.}~\bibnamefont {Polino}}, \bibinfo {author} {\bibfnamefont {M.}~\bibnamefont {Valeri}}, \bibinfo {author} {\bibfnamefont {N.}~\bibnamefont {Spagnolo}},\ and\ \bibinfo {author} {\bibfnamefont {F.}~\bibnamefont {Sciarrino}},\ }\bibfield  {title} {\bibinfo {title} {Benchmarking bayesian quantum estimation},\ }\href@noop {} {\bibfield  {journal} {\bibinfo  {journal} {Quantum Science and Technology}\ }\textbf {\bibinfo {volume} {9}},\ \bibinfo {pages} {035035} (\bibinfo {year} {2024})}\BibitemShut {NoStop}%
\bibitem [{\citenamefont {Salvia}\ \emph {et~al.}(2023)\citenamefont {Salvia}, \citenamefont {Mehboudi},\ and\ \citenamefont {Perarnau-Llobet}}]{salvia2023critical}%
  \BibitemOpen
  \bibfield  {author} {\bibinfo {author} {\bibfnamefont {R.}~\bibnamefont {Salvia}}, \bibinfo {author} {\bibfnamefont {M.}~\bibnamefont {Mehboudi}},\ and\ \bibinfo {author} {\bibfnamefont {M.}~\bibnamefont {Perarnau-Llobet}},\ }\bibfield  {title} {\bibinfo {title} {Critical quantum metrology assisted by real-time feedback control},\ }\href@noop {} {\bibfield  {journal} {\bibinfo  {journal} {Phys. Rev. Lett.}\ }\textbf {\bibinfo {volume} {130}},\ \bibinfo {pages} {240803} (\bibinfo {year} {2023})}\BibitemShut {NoStop}%
\bibitem [{\citenamefont {Van~Trees}(2004)}]{van2004detection}%
  \BibitemOpen
  \bibfield  {author} {\bibinfo {author} {\bibfnamefont {H.~L.}\ \bibnamefont {Van~Trees}},\ }\href@noop {} {\emph {\bibinfo {title} {Detection, estimation, and modulation theory, part I: detection, estimation, and linear modulation theory}}}\ (\bibinfo  {publisher} {John Wiley \& Sons},\ \bibinfo {year} {2004})\BibitemShut {NoStop}%
\bibitem [{\citenamefont {Gill}\ and\ \citenamefont {Levit}(1995)}]{gill1995applications}%
  \BibitemOpen
  \bibfield  {author} {\bibinfo {author} {\bibfnamefont {R.~D.}\ \bibnamefont {Gill}}\ and\ \bibinfo {author} {\bibfnamefont {B.~Y.}\ \bibnamefont {Levit}},\ }\bibfield  {title} {\bibinfo {title} {Applications of the van trees inequality: a bayesian cram{\'e}r-rao bound},\ }\href@noop {} {\bibfield  {journal} {\bibinfo  {journal} {Bernoulli}\ }\textbf {\bibinfo {volume} {1}},\ \bibinfo {pages} {59} (\bibinfo {year} {1995})}\BibitemShut {NoStop}%
\bibitem [{\citenamefont {Montenegro}\ \emph {et~al.}(2021)\citenamefont {Montenegro}, \citenamefont {Mishra},\ and\ \citenamefont {Bayat}}]{montenegro2021global}%
  \BibitemOpen
  \bibfield  {author} {\bibinfo {author} {\bibfnamefont {V.}~\bibnamefont {Montenegro}}, \bibinfo {author} {\bibfnamefont {U.}~\bibnamefont {Mishra}},\ and\ \bibinfo {author} {\bibfnamefont {A.}~\bibnamefont {Bayat}},\ }\bibfield  {title} {\bibinfo {title} {Global sensing and its impact for quantum many-body probes with criticality},\ }\href@noop {} {\bibfield  {journal} {\bibinfo  {journal} {Physical Review Letters}\ }\textbf {\bibinfo {volume} {126}},\ \bibinfo {pages} {200501} (\bibinfo {year} {2021})}\BibitemShut {NoStop}%
\bibitem [{\citenamefont {Serafini}(2017)}]{serafini2017quantum}%
  \BibitemOpen
  \bibfield  {author} {\bibinfo {author} {\bibfnamefont {A.}~\bibnamefont {Serafini}},\ }\href@noop {} {\emph {\bibinfo {title} {Quantum continuous variables: a primer of theoretical methods}}}\ (\bibinfo  {publisher} {CRC press},\ \bibinfo {year} {2017})\BibitemShut {NoStop}%
\bibitem [{\citenamefont {Ferraro}\ \emph {et~al.}(2005)\citenamefont {Ferraro}, \citenamefont {Olivares},\ and\ \citenamefont {Paris}}]{ferraro2005gaussian}%
  \BibitemOpen
  \bibfield  {author} {\bibinfo {author} {\bibfnamefont {A.}~\bibnamefont {Ferraro}}, \bibinfo {author} {\bibfnamefont {S.}~\bibnamefont {Olivares}},\ and\ \bibinfo {author} {\bibfnamefont {M.~G.}\ \bibnamefont {Paris}},\ }\bibfield  {title} {\bibinfo {title} {Gaussian states in continuous variable quantum information},\ }\href@noop {} {\bibfield  {journal} {\bibinfo  {journal} {arXiv preprint quant-ph/0503237}\ } (\bibinfo {year} {2005})}\BibitemShut {NoStop}%
\bibitem [{\citenamefont {Wang}\ \emph {et~al.}(2007)\citenamefont {Wang}, \citenamefont {Hiroshima}, \citenamefont {Tomita},\ and\ \citenamefont {Hayashi}}]{wang2007quantum}%
  \BibitemOpen
  \bibfield  {author} {\bibinfo {author} {\bibfnamefont {X.-B.}\ \bibnamefont {Wang}}, \bibinfo {author} {\bibfnamefont {T.}~\bibnamefont {Hiroshima}}, \bibinfo {author} {\bibfnamefont {A.}~\bibnamefont {Tomita}},\ and\ \bibinfo {author} {\bibfnamefont {M.}~\bibnamefont {Hayashi}},\ }\bibfield  {title} {\bibinfo {title} {Quantum information with gaussian states},\ }\href@noop {} {\bibfield  {journal} {\bibinfo  {journal} {Physics reports}\ }\textbf {\bibinfo {volume} {448}},\ \bibinfo {pages} {1} (\bibinfo {year} {2007})}\BibitemShut {NoStop}%
\bibitem [{\citenamefont {Weedbrook}\ \emph {et~al.}(2012)\citenamefont {Weedbrook}, \citenamefont {Pirandola}, \citenamefont {Garc{\'\i}a-Patr{\'o}n}, \citenamefont {Cerf}, \citenamefont {Ralph}, \citenamefont {Shapiro},\ and\ \citenamefont {Lloyd}}]{weedbrook2012gaussian}%
  \BibitemOpen
  \bibfield  {author} {\bibinfo {author} {\bibfnamefont {C.}~\bibnamefont {Weedbrook}}, \bibinfo {author} {\bibfnamefont {S.}~\bibnamefont {Pirandola}}, \bibinfo {author} {\bibfnamefont {R.}~\bibnamefont {Garc{\'\i}a-Patr{\'o}n}}, \bibinfo {author} {\bibfnamefont {N.~J.}\ \bibnamefont {Cerf}}, \bibinfo {author} {\bibfnamefont {T.~C.}\ \bibnamefont {Ralph}}, \bibinfo {author} {\bibfnamefont {J.~H.}\ \bibnamefont {Shapiro}},\ and\ \bibinfo {author} {\bibfnamefont {S.}~\bibnamefont {Lloyd}},\ }\bibfield  {title} {\bibinfo {title} {Gaussian quantum information},\ }\href@noop {} {\bibfield  {journal} {\bibinfo  {journal} {Reviews of Modern Physics}\ }\textbf {\bibinfo {volume} {84}},\ \bibinfo {pages} {621} (\bibinfo {year} {2012})}\BibitemShut {NoStop}%
\bibitem [{\citenamefont {Adesso}\ \emph {et~al.}(2014)\citenamefont {Adesso}, \citenamefont {Ragy},\ and\ \citenamefont {Lee}}]{adesso2014continuous}%
  \BibitemOpen
  \bibfield  {author} {\bibinfo {author} {\bibfnamefont {G.}~\bibnamefont {Adesso}}, \bibinfo {author} {\bibfnamefont {S.}~\bibnamefont {Ragy}},\ and\ \bibinfo {author} {\bibfnamefont {A.~R.}\ \bibnamefont {Lee}},\ }\bibfield  {title} {\bibinfo {title} {Continuous variable quantum information: Gaussian states and beyond},\ }\href@noop {} {\bibfield  {journal} {\bibinfo  {journal} {Open Systems \& Information Dynamics}\ }\textbf {\bibinfo {volume} {21}},\ \bibinfo {pages} {1440001} (\bibinfo {year} {2014})}\BibitemShut {NoStop}%
\bibitem [{\citenamefont {Serafini}\ \emph {et~al.}(2003)\citenamefont {Serafini}, \citenamefont {Illuminati},\ and\ \citenamefont {De~Siena}}]{serafini2003symplectic}%
  \BibitemOpen
  \bibfield  {author} {\bibinfo {author} {\bibfnamefont {A.}~\bibnamefont {Serafini}}, \bibinfo {author} {\bibfnamefont {F.}~\bibnamefont {Illuminati}},\ and\ \bibinfo {author} {\bibfnamefont {S.}~\bibnamefont {De~Siena}},\ }\bibfield  {title} {\bibinfo {title} {Symplectic invariants, entropic measures and correlations of gaussian states},\ }\href@noop {} {\bibfield  {journal} {\bibinfo  {journal} {Journal of Physics B: Atomic, Molecular and Optical Physics}\ }\textbf {\bibinfo {volume} {37}},\ \bibinfo {pages} {L21} (\bibinfo {year} {2003})}\BibitemShut {NoStop}%
\bibitem [{\citenamefont {Giedke}\ and\ \citenamefont {Cirac}(2002)}]{giedke2002characterization}%
  \BibitemOpen
  \bibfield  {author} {\bibinfo {author} {\bibfnamefont {G.}~\bibnamefont {Giedke}}\ and\ \bibinfo {author} {\bibfnamefont {J.~I.}\ \bibnamefont {Cirac}},\ }\bibfield  {title} {\bibinfo {title} {Characterization of gaussian operations and distillation of gaussian states},\ }\href@noop {} {\bibfield  {journal} {\bibinfo  {journal} {Physical Review A}\ }\textbf {\bibinfo {volume} {66}},\ \bibinfo {pages} {032316} (\bibinfo {year} {2002})}\BibitemShut {NoStop}%
\bibitem [{\citenamefont {Filip}(2013)}]{filip2013distillation}%
  \BibitemOpen
  \bibfield  {author} {\bibinfo {author} {\bibfnamefont {R.}~\bibnamefont {Filip}},\ }\bibfield  {title} {\bibinfo {title} {Distillation of quantum squeezing},\ }\href@noop {} {\bibfield  {journal} {\bibinfo  {journal} {Physical Review A—Atomic, Molecular, and Optical Physics}\ }\textbf {\bibinfo {volume} {88}},\ \bibinfo {pages} {063837} (\bibinfo {year} {2013})}\BibitemShut {NoStop}%
\bibitem [{\citenamefont {Rugar}\ and\ \citenamefont {Gr{\"u}tter}(1991)}]{rugar1991mechanical}%
  \BibitemOpen
  \bibfield  {author} {\bibinfo {author} {\bibfnamefont {D.}~\bibnamefont {Rugar}}\ and\ \bibinfo {author} {\bibfnamefont {P.}~\bibnamefont {Gr{\"u}tter}},\ }\bibfield  {title} {\bibinfo {title} {Mechanical parametric amplification and thermomechanical noise squeezing},\ }\href@noop {} {\bibfield  {journal} {\bibinfo  {journal} {Physical Review Letters}\ }\textbf {\bibinfo {volume} {67}},\ \bibinfo {pages} {699} (\bibinfo {year} {1991})}\BibitemShut {NoStop}%
\bibitem [{\citenamefont {Olivares}\ and\ \citenamefont {Paris}(2007)}]{olivares2007optimized}%
  \BibitemOpen
  \bibfield  {author} {\bibinfo {author} {\bibfnamefont {S.}~\bibnamefont {Olivares}}\ and\ \bibinfo {author} {\bibfnamefont {M.~G.~A.}\ \bibnamefont {Paris}},\ }\bibfield  {title} {\bibinfo {title} {Optimized interferometry with gaussian states},\ }\href@noop {} {\bibfield  {journal} {\bibinfo  {journal} {Optics and Spectroscopy}\ }\textbf {\bibinfo {volume} {103}},\ \bibinfo {pages} {231} (\bibinfo {year} {2007})}\BibitemShut {NoStop}%
\bibitem [{\citenamefont {Sparaciari}\ \emph {et~al.}(2015)\citenamefont {Sparaciari}, \citenamefont {Olivares},\ and\ \citenamefont {Paris}}]{sparaciari2015bounds}%
  \BibitemOpen
  \bibfield  {author} {\bibinfo {author} {\bibfnamefont {C.}~\bibnamefont {Sparaciari}}, \bibinfo {author} {\bibfnamefont {S.}~\bibnamefont {Olivares}},\ and\ \bibinfo {author} {\bibfnamefont {M.~G.~A.}\ \bibnamefont {Paris}},\ }\bibfield  {title} {\bibinfo {title} {Bounds to precision for quantum interferometry with gaussian states and operations},\ }\href@noop {} {\bibfield  {journal} {\bibinfo  {journal} {JOSA B}\ }\textbf {\bibinfo {volume} {32}},\ \bibinfo {pages} {1354} (\bibinfo {year} {2015})}\BibitemShut {NoStop}%
\bibitem [{\citenamefont {Adesso}(2014)}]{adesso2014gaussian}%
  \BibitemOpen
  \bibfield  {author} {\bibinfo {author} {\bibfnamefont {G.}~\bibnamefont {Adesso}},\ }\bibfield  {title} {\bibinfo {title} {Gaussian interferometric power},\ }\href@noop {} {\bibfield  {journal} {\bibinfo  {journal} {Physical Review A}\ }\textbf {\bibinfo {volume} {90}},\ \bibinfo {pages} {022321} (\bibinfo {year} {2014})}\BibitemShut {NoStop}%
\bibitem [{\citenamefont {Sanz}\ \emph {et~al.}(2017)\citenamefont {Sanz}, \citenamefont {Las~Heras}, \citenamefont {Garc{\'\i}a-Ripoll}, \citenamefont {Solano},\ and\ \citenamefont {Di~Candia}}]{sanz2017quantum}%
  \BibitemOpen
  \bibfield  {author} {\bibinfo {author} {\bibfnamefont {M.}~\bibnamefont {Sanz}}, \bibinfo {author} {\bibfnamefont {U.}~\bibnamefont {Las~Heras}}, \bibinfo {author} {\bibfnamefont {J.~J.}\ \bibnamefont {Garc{\'\i}a-Ripoll}}, \bibinfo {author} {\bibfnamefont {E.}~\bibnamefont {Solano}},\ and\ \bibinfo {author} {\bibfnamefont {R.}~\bibnamefont {Di~Candia}},\ }\bibfield  {title} {\bibinfo {title} {Quantum estimation methods for quantum illumination},\ }\href@noop {} {\bibfield  {journal} {\bibinfo  {journal} {Physical review letters}\ }\textbf {\bibinfo {volume} {118}},\ \bibinfo {pages} {070803} (\bibinfo {year} {2017})}\BibitemShut {NoStop}%
\bibitem [{\citenamefont {Friis}\ \emph {et~al.}(2015)\citenamefont {Friis}, \citenamefont {Skotiniotis}, \citenamefont {Fuentes},\ and\ \citenamefont {D{\"u}r}}]{friis2015heisenberg}%
  \BibitemOpen
  \bibfield  {author} {\bibinfo {author} {\bibfnamefont {N.}~\bibnamefont {Friis}}, \bibinfo {author} {\bibfnamefont {M.}~\bibnamefont {Skotiniotis}}, \bibinfo {author} {\bibfnamefont {I.}~\bibnamefont {Fuentes}},\ and\ \bibinfo {author} {\bibfnamefont {W.}~\bibnamefont {D{\"u}r}},\ }\bibfield  {title} {\bibinfo {title} {Heisenberg scaling in gaussian quantum metrology},\ }\href@noop {} {\bibfield  {journal} {\bibinfo  {journal} {Physical Review A}\ }\textbf {\bibinfo {volume} {92}},\ \bibinfo {pages} {022106} (\bibinfo {year} {2015})}\BibitemShut {NoStop}%
\bibitem [{\citenamefont {Ruppert}\ and\ \citenamefont {Filip}(2017)}]{ruppert2017light}%
  \BibitemOpen
  \bibfield  {author} {\bibinfo {author} {\bibfnamefont {L.}~\bibnamefont {Ruppert}}\ and\ \bibinfo {author} {\bibfnamefont {R.}~\bibnamefont {Filip}},\ }\bibfield  {title} {\bibinfo {title} {Light-matter quantum interferometry with homodyne detection},\ }\href@noop {} {\bibfield  {journal} {\bibinfo  {journal} {Optics Express}\ }\textbf {\bibinfo {volume} {25}},\ \bibinfo {pages} {15456} (\bibinfo {year} {2017})}\BibitemShut {NoStop}%
\bibitem [{\citenamefont {Matsubara}\ \emph {et~al.}(2019)\citenamefont {Matsubara}, \citenamefont {Facchi}, \citenamefont {Giovannetti},\ and\ \citenamefont {Yuasa}}]{matsubara2019optimal}%
  \BibitemOpen
  \bibfield  {author} {\bibinfo {author} {\bibfnamefont {T.}~\bibnamefont {Matsubara}}, \bibinfo {author} {\bibfnamefont {P.}~\bibnamefont {Facchi}}, \bibinfo {author} {\bibfnamefont {V.}~\bibnamefont {Giovannetti}},\ and\ \bibinfo {author} {\bibfnamefont {K.}~\bibnamefont {Yuasa}},\ }\bibfield  {title} {\bibinfo {title} {Optimal gaussian metrology for generic multimode interferometric circuit},\ }\href@noop {} {\bibfield  {journal} {\bibinfo  {journal} {New Journal of Physics}\ }\textbf {\bibinfo {volume} {21}},\ \bibinfo {pages} {033014} (\bibinfo {year} {2019})}\BibitemShut {NoStop}%
\bibitem [{\citenamefont {Motes}\ \emph {et~al.}(2015)\citenamefont {Motes}, \citenamefont {Olson}, \citenamefont {Rabeaux}, \citenamefont {Dowling}, \citenamefont {Olson},\ and\ \citenamefont {Rohde}}]{motes2015linear}%
  \BibitemOpen
  \bibfield  {author} {\bibinfo {author} {\bibfnamefont {K.~R.}\ \bibnamefont {Motes}}, \bibinfo {author} {\bibfnamefont {J.~P.}\ \bibnamefont {Olson}}, \bibinfo {author} {\bibfnamefont {E.~J.}\ \bibnamefont {Rabeaux}}, \bibinfo {author} {\bibfnamefont {J.~P.}\ \bibnamefont {Dowling}}, \bibinfo {author} {\bibfnamefont {S.~J.}\ \bibnamefont {Olson}},\ and\ \bibinfo {author} {\bibfnamefont {P.~P.}\ \bibnamefont {Rohde}},\ }\bibfield  {title} {\bibinfo {title} {Linear optical quantum metrology with single photons: exploiting spontaneously generated entanglement to beat the shot-noise limit},\ }\href@noop {} {\bibfield  {journal} {\bibinfo  {journal} {Physical review letters}\ }\textbf {\bibinfo {volume} {114}},\ \bibinfo {pages} {170802} (\bibinfo {year} {2015})}\BibitemShut {NoStop}%
\bibitem [{\citenamefont {Sparaciari}\ \emph {et~al.}(2016)\citenamefont {Sparaciari}, \citenamefont {Olivares},\ and\ \citenamefont {Paris}}]{sparaciari2016gaussian}%
  \BibitemOpen
  \bibfield  {author} {\bibinfo {author} {\bibfnamefont {C.}~\bibnamefont {Sparaciari}}, \bibinfo {author} {\bibfnamefont {S.}~\bibnamefont {Olivares}},\ and\ \bibinfo {author} {\bibfnamefont {M.~G.~A.}\ \bibnamefont {Paris}},\ }\bibfield  {title} {\bibinfo {title} {Gaussian-state interferometry with passive and active elements},\ }\href@noop {} {\bibfield  {journal} {\bibinfo  {journal} {Physical Review A}\ }\textbf {\bibinfo {volume} {93}},\ \bibinfo {pages} {023810} (\bibinfo {year} {2016})}\BibitemShut {NoStop}%
\bibitem [{\citenamefont {Nichols}\ \emph {et~al.}(2018)\citenamefont {Nichols}, \citenamefont {Liuzzo-Scorpo}, \citenamefont {Knott},\ and\ \citenamefont {Adesso}}]{nichols2018multiparameter}%
  \BibitemOpen
  \bibfield  {author} {\bibinfo {author} {\bibfnamefont {R.}~\bibnamefont {Nichols}}, \bibinfo {author} {\bibfnamefont {P.}~\bibnamefont {Liuzzo-Scorpo}}, \bibinfo {author} {\bibfnamefont {P.~A.}\ \bibnamefont {Knott}},\ and\ \bibinfo {author} {\bibfnamefont {G.}~\bibnamefont {Adesso}},\ }\bibfield  {title} {\bibinfo {title} {Multiparameter gaussian quantum metrology},\ }\href@noop {} {\bibfield  {journal} {\bibinfo  {journal} {Physical Review A}\ }\textbf {\bibinfo {volume} {98}},\ \bibinfo {pages} {012114} (\bibinfo {year} {2018})}\BibitemShut {NoStop}%
\bibitem [{\citenamefont {Huang}\ and\ \citenamefont {Moore}(2008)}]{huang2008optimized}%
  \BibitemOpen
  \bibfield  {author} {\bibinfo {author} {\bibfnamefont {Y.}~\bibnamefont {Huang}}\ and\ \bibinfo {author} {\bibfnamefont {M.}~\bibnamefont {Moore}},\ }\bibfield  {title} {\bibinfo {title} {Optimized double-well quantum interferometry with gaussian squeezed states},\ }\href@noop {} {\bibfield  {journal} {\bibinfo  {journal} {Physical review letters}\ }\textbf {\bibinfo {volume} {100}},\ \bibinfo {pages} {250406} (\bibinfo {year} {2008})}\BibitemShut {NoStop}%
\bibitem [{\citenamefont {Yadin}\ \emph {et~al.}(2018)\citenamefont {Yadin}, \citenamefont {Binder}, \citenamefont {Thompson}, \citenamefont {Narasimhachar}, \citenamefont {Gu},\ and\ \citenamefont {Kim}}]{yadin2018operational}%
  \BibitemOpen
  \bibfield  {author} {\bibinfo {author} {\bibfnamefont {B.}~\bibnamefont {Yadin}}, \bibinfo {author} {\bibfnamefont {F.~C.}\ \bibnamefont {Binder}}, \bibinfo {author} {\bibfnamefont {J.}~\bibnamefont {Thompson}}, \bibinfo {author} {\bibfnamefont {V.}~\bibnamefont {Narasimhachar}}, \bibinfo {author} {\bibfnamefont {M.}~\bibnamefont {Gu}},\ and\ \bibinfo {author} {\bibfnamefont {M.}~\bibnamefont {Kim}},\ }\bibfield  {title} {\bibinfo {title} {Operational resource theory of continuous-variable nonclassicality},\ }\href@noop {} {\bibfield  {journal} {\bibinfo  {journal} {Physical Review X}\ }\textbf {\bibinfo {volume} {8}},\ \bibinfo {pages} {041038} (\bibinfo {year} {2018})}\BibitemShut {NoStop}%
\bibitem [{\citenamefont {Laurenza}\ \emph {et~al.}(2018)\citenamefont {Laurenza}, \citenamefont {Lupo}, \citenamefont {Spedalieri}, \citenamefont {Braunstein},\ and\ \citenamefont {Pirandola}}]{laurenza2018channel}%
  \BibitemOpen
  \bibfield  {author} {\bibinfo {author} {\bibfnamefont {R.}~\bibnamefont {Laurenza}}, \bibinfo {author} {\bibfnamefont {C.}~\bibnamefont {Lupo}}, \bibinfo {author} {\bibfnamefont {G.}~\bibnamefont {Spedalieri}}, \bibinfo {author} {\bibfnamefont {S.~L.}\ \bibnamefont {Braunstein}},\ and\ \bibinfo {author} {\bibfnamefont {S.}~\bibnamefont {Pirandola}},\ }\bibfield  {title} {\bibinfo {title} {Channel simulation in quantum metrology},\ }\href@noop {} {\bibfield  {journal} {\bibinfo  {journal} {Quantum Measurements and Quantum Metrology}\ }\textbf {\bibinfo {volume} {5}},\ \bibinfo {pages} {1} (\bibinfo {year} {2018})}\BibitemShut {NoStop}%
\bibitem [{\citenamefont {Oh}\ \emph {et~al.}(2019{\natexlab{a}})\citenamefont {Oh}, \citenamefont {Lee}, \citenamefont {Banchi}, \citenamefont {Lee}, \citenamefont {Rockstuhl},\ and\ \citenamefont {Jeong}}]{oh2019optimalpra}%
  \BibitemOpen
  \bibfield  {author} {\bibinfo {author} {\bibfnamefont {C.}~\bibnamefont {Oh}}, \bibinfo {author} {\bibfnamefont {C.}~\bibnamefont {Lee}}, \bibinfo {author} {\bibfnamefont {L.}~\bibnamefont {Banchi}}, \bibinfo {author} {\bibfnamefont {S.-Y.}\ \bibnamefont {Lee}}, \bibinfo {author} {\bibfnamefont {C.}~\bibnamefont {Rockstuhl}},\ and\ \bibinfo {author} {\bibfnamefont {H.}~\bibnamefont {Jeong}},\ }\bibfield  {title} {\bibinfo {title} {Optimal measurements for quantum fidelity between gaussian states and its relevance to quantum metrology},\ }\href@noop {} {\bibfield  {journal} {\bibinfo  {journal} {Physical Review A}\ }\textbf {\bibinfo {volume} {100}},\ \bibinfo {pages} {012323} (\bibinfo {year} {2019}{\natexlab{a}})}\BibitemShut {NoStop}%
\bibitem [{\citenamefont {Oh}\ \emph {et~al.}(2019{\natexlab{b}})\citenamefont {Oh}, \citenamefont {Lee}, \citenamefont {Rockstuhl}, \citenamefont {Jeong}, \citenamefont {Kim}, \citenamefont {Nha},\ and\ \citenamefont {Lee}}]{oh2019optimal}%
  \BibitemOpen
  \bibfield  {author} {\bibinfo {author} {\bibfnamefont {C.}~\bibnamefont {Oh}}, \bibinfo {author} {\bibfnamefont {C.}~\bibnamefont {Lee}}, \bibinfo {author} {\bibfnamefont {C.}~\bibnamefont {Rockstuhl}}, \bibinfo {author} {\bibfnamefont {H.}~\bibnamefont {Jeong}}, \bibinfo {author} {\bibfnamefont {J.}~\bibnamefont {Kim}}, \bibinfo {author} {\bibfnamefont {H.}~\bibnamefont {Nha}},\ and\ \bibinfo {author} {\bibfnamefont {S.-Y.}\ \bibnamefont {Lee}},\ }\bibfield  {title} {\bibinfo {title} {Optimal gaussian measurements for phase estimation in single-mode gaussian metrology},\ }\href@noop {} {\bibfield  {journal} {\bibinfo  {journal} {npj Quantum Information}\ }\textbf {\bibinfo {volume} {5}},\ \bibinfo {pages} {10} (\bibinfo {year} {2019}{\natexlab{b}})}\BibitemShut {NoStop}%
\bibitem [{\citenamefont {Bakmou}\ \emph {et~al.}(2019)\citenamefont {Bakmou}, \citenamefont {Slaoui}, \citenamefont {Daoud},\ and\ \citenamefont {Ahl~Laamara}}]{bakmou2019quantum}%
  \BibitemOpen
  \bibfield  {author} {\bibinfo {author} {\bibfnamefont {L.}~\bibnamefont {Bakmou}}, \bibinfo {author} {\bibfnamefont {A.}~\bibnamefont {Slaoui}}, \bibinfo {author} {\bibfnamefont {M.}~\bibnamefont {Daoud}},\ and\ \bibinfo {author} {\bibfnamefont {R.}~\bibnamefont {Ahl~Laamara}},\ }\bibfield  {title} {\bibinfo {title} {Quantum fisher information matrix in heisenberg xy model},\ }\href@noop {} {\bibfield  {journal} {\bibinfo  {journal} {Quantum Information Processing}\ }\textbf {\bibinfo {volume} {18}},\ \bibinfo {pages} {1} (\bibinfo {year} {2019})}\BibitemShut {NoStop}%
\bibitem [{\citenamefont {Sorelli}\ \emph {et~al.}(2024)\citenamefont {Sorelli}, \citenamefont {Gessner}, \citenamefont {Treps},\ and\ \citenamefont {Walschaers}}]{sorelli2024gaussian}%
  \BibitemOpen
  \bibfield  {author} {\bibinfo {author} {\bibfnamefont {G.}~\bibnamefont {Sorelli}}, \bibinfo {author} {\bibfnamefont {M.}~\bibnamefont {Gessner}}, \bibinfo {author} {\bibfnamefont {N.}~\bibnamefont {Treps}},\ and\ \bibinfo {author} {\bibfnamefont {M.}~\bibnamefont {Walschaers}},\ }\bibfield  {title} {\bibinfo {title} {Gaussian quantum metrology for mode-encoded parameters},\ }\href@noop {} {\bibfield  {journal} {\bibinfo  {journal} {New Journal of Physics}\ }\textbf {\bibinfo {volume} {26}},\ \bibinfo {pages} {073022} (\bibinfo {year} {2024})}\BibitemShut {NoStop}%
\bibitem [{\citenamefont {Porto}\ \emph {et~al.}(2024)\citenamefont {Porto}, \citenamefont {Marinho}, \citenamefont {Dieguez}, \citenamefont {da~Paz},\ and\ \citenamefont {Vieira}}]{porto2024enhancing}%
  \BibitemOpen
  \bibfield  {author} {\bibinfo {author} {\bibfnamefont {J.~C.}\ \bibnamefont {Porto}}, \bibinfo {author} {\bibfnamefont {L.~S.}\ \bibnamefont {Marinho}}, \bibinfo {author} {\bibfnamefont {P.~R.}\ \bibnamefont {Dieguez}}, \bibinfo {author} {\bibfnamefont {I.~G.}\ \bibnamefont {da~Paz}},\ and\ \bibinfo {author} {\bibfnamefont {C.~H.}\ \bibnamefont {Vieira}},\ }\bibfield  {title} {\bibinfo {title} {Enhancing gaussian quantum metrology with position-momentum correlations},\ }\href@noop {} {\bibfield  {journal} {\bibinfo  {journal} {arXiv preprint arXiv:2408.13060}\ } (\bibinfo {year} {2024})}\BibitemShut {NoStop}%
\bibitem [{\citenamefont {Liao}\ and\ \citenamefont {Fu}(2022)}]{liao2022quantum}%
  \BibitemOpen
  \bibfield  {author} {\bibinfo {author} {\bibfnamefont {X.-J.}\ \bibnamefont {Liao}}\ and\ \bibinfo {author} {\bibfnamefont {Y.-Q.}\ \bibnamefont {Fu}},\ }\bibfield  {title} {\bibinfo {title} {Quantum metrology with multimode gaussian states of multiple point sources},\ }\href@noop {} {\bibfield  {journal} {\bibinfo  {journal} {Physical Review A}\ }\textbf {\bibinfo {volume} {106}},\ \bibinfo {pages} {022602} (\bibinfo {year} {2022})}\BibitemShut {NoStop}%
\bibitem [{\citenamefont {Hayashi}\ \emph {et~al.}(2022)\citenamefont {Hayashi}, \citenamefont {Liu},\ and\ \citenamefont {Yuan}}]{hayashi2022global}%
  \BibitemOpen
  \bibfield  {author} {\bibinfo {author} {\bibfnamefont {M.}~\bibnamefont {Hayashi}}, \bibinfo {author} {\bibfnamefont {Z.-W.}\ \bibnamefont {Liu}},\ and\ \bibinfo {author} {\bibfnamefont {H.}~\bibnamefont {Yuan}},\ }\bibfield  {title} {\bibinfo {title} {Global heisenberg scaling in noisy and practical phase estimation},\ }\href@noop {} {\bibfield  {journal} {\bibinfo  {journal} {Quantum Science and Technology}\ }\textbf {\bibinfo {volume} {7}},\ \bibinfo {pages} {025030} (\bibinfo {year} {2022})}\BibitemShut {NoStop}%
\bibitem [{\citenamefont {Zanardi}\ and\ \citenamefont {Paunkovi{\'c}}(2006)}]{zanardi2006ground}%
  \BibitemOpen
  \bibfield  {author} {\bibinfo {author} {\bibfnamefont {P.}~\bibnamefont {Zanardi}}\ and\ \bibinfo {author} {\bibfnamefont {N.}~\bibnamefont {Paunkovi{\'c}}},\ }\bibfield  {title} {\bibinfo {title} {Ground state overlap and quantum phase transitions},\ }\href@noop {} {\bibfield  {journal} {\bibinfo  {journal} {Physical Review E}\ }\textbf {\bibinfo {volume} {74}},\ \bibinfo {pages} {031123} (\bibinfo {year} {2006})}\BibitemShut {NoStop}%
\bibitem [{\citenamefont {Pirandola}\ \emph {et~al.}(2018)\citenamefont {Pirandola}, \citenamefont {Bardhan}, \citenamefont {Gehring}, \citenamefont {Weedbrook},\ and\ \citenamefont {Lloyd}}]{pirandola2018advances}%
  \BibitemOpen
  \bibfield  {author} {\bibinfo {author} {\bibfnamefont {S.}~\bibnamefont {Pirandola}}, \bibinfo {author} {\bibfnamefont {B.~R.}\ \bibnamefont {Bardhan}}, \bibinfo {author} {\bibfnamefont {T.}~\bibnamefont {Gehring}}, \bibinfo {author} {\bibfnamefont {C.}~\bibnamefont {Weedbrook}},\ and\ \bibinfo {author} {\bibfnamefont {S.}~\bibnamefont {Lloyd}},\ }\bibfield  {title} {\bibinfo {title} {Advances in photonic quantum sensing},\ }\href@noop {} {\bibfield  {journal} {\bibinfo  {journal} {Nature Photonics}\ }\textbf {\bibinfo {volume} {12}},\ \bibinfo {pages} {724} (\bibinfo {year} {2018})}\BibitemShut {NoStop}%
\bibitem [{\citenamefont {Dowling}\ and\ \citenamefont {Seshadreesan}(2015)}]{dowling2015quantum}%
  \BibitemOpen
  \bibfield  {author} {\bibinfo {author} {\bibfnamefont {J.~P.}\ \bibnamefont {Dowling}}\ and\ \bibinfo {author} {\bibfnamefont {K.~P.}\ \bibnamefont {Seshadreesan}},\ }\bibfield  {title} {\bibinfo {title} {Quantum optical technologies for metrology, sensing, and imaging},\ }\href@noop {} {\bibfield  {journal} {\bibinfo  {journal} {Journal of Lightwave Technology}\ }\textbf {\bibinfo {volume} {33}},\ \bibinfo {pages} {2359} (\bibinfo {year} {2015})}\BibitemShut {NoStop}%
\bibitem [{\citenamefont {Oh}\ \emph {et~al.}(2017)\citenamefont {Oh}, \citenamefont {Lee}, \citenamefont {Nha},\ and\ \citenamefont {Jeong}}]{oh2017practical}%
  \BibitemOpen
  \bibfield  {author} {\bibinfo {author} {\bibfnamefont {C.}~\bibnamefont {Oh}}, \bibinfo {author} {\bibfnamefont {S.-Y.}\ \bibnamefont {Lee}}, \bibinfo {author} {\bibfnamefont {H.}~\bibnamefont {Nha}},\ and\ \bibinfo {author} {\bibfnamefont {H.}~\bibnamefont {Jeong}},\ }\bibfield  {title} {\bibinfo {title} {Practical resources and measurements for lossy optical quantum metrology},\ }\href@noop {} {\bibfield  {journal} {\bibinfo  {journal} {Physical Review A}\ }\textbf {\bibinfo {volume} {96}},\ \bibinfo {pages} {062304} (\bibinfo {year} {2017})}\BibitemShut {NoStop}%
\bibitem [{\citenamefont {Raffaelli}\ \emph {et~al.}(2018)\citenamefont {Raffaelli}, \citenamefont {Ferranti}, \citenamefont {Mahler}, \citenamefont {Sibson}, \citenamefont {Kennard}, \citenamefont {Santamato}, \citenamefont {Sinclair}, \citenamefont {Bonneau}, \citenamefont {Thompson},\ and\ \citenamefont {Matthews}}]{raffaelli2018homodyne}%
  \BibitemOpen
  \bibfield  {author} {\bibinfo {author} {\bibfnamefont {F.}~\bibnamefont {Raffaelli}}, \bibinfo {author} {\bibfnamefont {G.}~\bibnamefont {Ferranti}}, \bibinfo {author} {\bibfnamefont {D.~H.}\ \bibnamefont {Mahler}}, \bibinfo {author} {\bibfnamefont {P.}~\bibnamefont {Sibson}}, \bibinfo {author} {\bibfnamefont {J.~E.}\ \bibnamefont {Kennard}}, \bibinfo {author} {\bibfnamefont {A.}~\bibnamefont {Santamato}}, \bibinfo {author} {\bibfnamefont {G.}~\bibnamefont {Sinclair}}, \bibinfo {author} {\bibfnamefont {D.}~\bibnamefont {Bonneau}}, \bibinfo {author} {\bibfnamefont {M.~G.}\ \bibnamefont {Thompson}},\ and\ \bibinfo {author} {\bibfnamefont {J.~C.}\ \bibnamefont {Matthews}},\ }\bibfield  {title} {\bibinfo {title} {A homodyne detector integrated onto a photonic chip for measuring quantum states and generating random numbers},\ }\href@noop {} {\bibfield  {journal} {\bibinfo  {journal} {Quantum Science and Technology}\ }\textbf {\bibinfo {volume} {3}},\ \bibinfo {pages} {025003} (\bibinfo {year}
  {2018})}\BibitemShut {NoStop}%
\bibitem [{\citenamefont {Wang}\ \emph {et~al.}(2019)\citenamefont {Wang}, \citenamefont {Wu}, \citenamefont {Ma}, \citenamefont {Cai}, \citenamefont {Hu}, \citenamefont {Mu}, \citenamefont {Xu}, \citenamefont {Chen}, \citenamefont {Wang}, \citenamefont {Song} \emph {et~al.}}]{wang2019heisenberg}%
  \BibitemOpen
  \bibfield  {author} {\bibinfo {author} {\bibfnamefont {W.}~\bibnamefont {Wang}}, \bibinfo {author} {\bibfnamefont {Y.}~\bibnamefont {Wu}}, \bibinfo {author} {\bibfnamefont {Y.}~\bibnamefont {Ma}}, \bibinfo {author} {\bibfnamefont {W.}~\bibnamefont {Cai}}, \bibinfo {author} {\bibfnamefont {L.}~\bibnamefont {Hu}}, \bibinfo {author} {\bibfnamefont {X.}~\bibnamefont {Mu}}, \bibinfo {author} {\bibfnamefont {Y.}~\bibnamefont {Xu}}, \bibinfo {author} {\bibfnamefont {Z.-J.}\ \bibnamefont {Chen}}, \bibinfo {author} {\bibfnamefont {H.}~\bibnamefont {Wang}}, \bibinfo {author} {\bibfnamefont {Y.}~\bibnamefont {Song}}, \emph {et~al.},\ }\bibfield  {title} {\bibinfo {title} {Heisenberg-limited single-mode quantum metrology in a superconducting circuit},\ }\href@noop {} {\bibfield  {journal} {\bibinfo  {journal} {Nature communications}\ }\textbf {\bibinfo {volume} {10}},\ \bibinfo {pages} {4382} (\bibinfo {year} {2019})}\BibitemShut {NoStop}%
\bibitem [{\citenamefont {Wolf}\ \emph {et~al.}(2019)\citenamefont {Wolf}, \citenamefont {Shi}, \citenamefont {Heip}, \citenamefont {Gessner}, \citenamefont {Pezz{\`e}}, \citenamefont {Smerzi}, \citenamefont {Schulte}, \citenamefont {Hammerer},\ and\ \citenamefont {Schmidt}}]{wolf2019motional}%
  \BibitemOpen
  \bibfield  {author} {\bibinfo {author} {\bibfnamefont {F.}~\bibnamefont {Wolf}}, \bibinfo {author} {\bibfnamefont {C.}~\bibnamefont {Shi}}, \bibinfo {author} {\bibfnamefont {J.~C.}\ \bibnamefont {Heip}}, \bibinfo {author} {\bibfnamefont {M.}~\bibnamefont {Gessner}}, \bibinfo {author} {\bibfnamefont {L.}~\bibnamefont {Pezz{\`e}}}, \bibinfo {author} {\bibfnamefont {A.}~\bibnamefont {Smerzi}}, \bibinfo {author} {\bibfnamefont {M.}~\bibnamefont {Schulte}}, \bibinfo {author} {\bibfnamefont {K.}~\bibnamefont {Hammerer}},\ and\ \bibinfo {author} {\bibfnamefont {P.~O.}\ \bibnamefont {Schmidt}},\ }\bibfield  {title} {\bibinfo {title} {Motional fock states for quantum-enhanced amplitude and phase measurements with trapped ions},\ }\href@noop {} {\bibfield  {journal} {\bibinfo  {journal} {Nature communications}\ }\textbf {\bibinfo {volume} {10}},\ \bibinfo {pages} {2929} (\bibinfo {year} {2019})}\BibitemShut {NoStop}%
\bibitem [{\citenamefont {McCormick}\ \emph {et~al.}(2019)\citenamefont {McCormick}, \citenamefont {Keller}, \citenamefont {Burd}, \citenamefont {Wineland}, \citenamefont {Wilson},\ and\ \citenamefont {Leibfried}}]{mccormick2019quantum}%
  \BibitemOpen
  \bibfield  {author} {\bibinfo {author} {\bibfnamefont {K.~C.}\ \bibnamefont {McCormick}}, \bibinfo {author} {\bibfnamefont {J.}~\bibnamefont {Keller}}, \bibinfo {author} {\bibfnamefont {S.~C.}\ \bibnamefont {Burd}}, \bibinfo {author} {\bibfnamefont {D.~J.}\ \bibnamefont {Wineland}}, \bibinfo {author} {\bibfnamefont {A.~C.}\ \bibnamefont {Wilson}},\ and\ \bibinfo {author} {\bibfnamefont {D.}~\bibnamefont {Leibfried}},\ }\bibfield  {title} {\bibinfo {title} {Quantum-enhanced sensing of a single-ion mechanical oscillator},\ }\href@noop {} {\bibfield  {journal} {\bibinfo  {journal} {Nature}\ }\textbf {\bibinfo {volume} {572}},\ \bibinfo {pages} {86} (\bibinfo {year} {2019})}\BibitemShut {NoStop}%
\bibitem [{\citenamefont {Pan}\ \emph {et~al.}(2024)\citenamefont {Pan}, \citenamefont {Krisnanda}, \citenamefont {Duina}, \citenamefont {Park}, \citenamefont {Song}, \citenamefont {Fontaine}, \citenamefont {Copetudo}, \citenamefont {Filip},\ and\ \citenamefont {Gao}}]{pan2024realisation}%
  \BibitemOpen
  \bibfield  {author} {\bibinfo {author} {\bibfnamefont {X.}~\bibnamefont {Pan}}, \bibinfo {author} {\bibfnamefont {T.}~\bibnamefont {Krisnanda}}, \bibinfo {author} {\bibfnamefont {A.}~\bibnamefont {Duina}}, \bibinfo {author} {\bibfnamefont {K.}~\bibnamefont {Park}}, \bibinfo {author} {\bibfnamefont {P.}~\bibnamefont {Song}}, \bibinfo {author} {\bibfnamefont {C.~Y.}\ \bibnamefont {Fontaine}}, \bibinfo {author} {\bibfnamefont {A.}~\bibnamefont {Copetudo}}, \bibinfo {author} {\bibfnamefont {R.}~\bibnamefont {Filip}},\ and\ \bibinfo {author} {\bibfnamefont {Y.~Y.}\ \bibnamefont {Gao}},\ }\bibfield  {title} {\bibinfo {title} {Realisation of versatile and effective quantum metrology using a single bosonic mode},\ }\href@noop {} {\bibfield  {journal} {\bibinfo  {journal} {arXiv preprint arXiv:2403.14967}\ } (\bibinfo {year} {2024})}\BibitemShut {NoStop}%
\bibitem [{\citenamefont {Fisher}(1922)}]{fisher1922mathematical}%
  \BibitemOpen
  \bibfield  {author} {\bibinfo {author} {\bibfnamefont {R.~A.}\ \bibnamefont {Fisher}},\ }\bibfield  {title} {\bibinfo {title} {On the mathematical foundations of theoretical statistics},\ }\href@noop {} {\bibfield  {journal} {\bibinfo  {journal} {Philosophical transactions of the Royal Society of London. Series A, containing papers of a mathematical or physical character}\ }\textbf {\bibinfo {volume} {222}},\ \bibinfo {pages} {309} (\bibinfo {year} {1922})}\BibitemShut {NoStop}%
\bibitem [{\citenamefont {Ko\l{}ody\ifmmode~\acute{n}\else \'{n}\fi{}ski}\ and\ \citenamefont {Demkowicz-Dobrza\ifmmode~\acute{n}\else \'{n}\fi{}ski}(2010)}]{kolodynski2010phase}%
  \BibitemOpen
  \bibfield  {author} {\bibinfo {author} {\bibfnamefont {J.}~\bibnamefont {Ko\l{}ody\ifmmode~\acute{n}\else \'{n}\fi{}ski}}\ and\ \bibinfo {author} {\bibfnamefont {R.}~\bibnamefont {Demkowicz-Dobrza\ifmmode~\acute{n}\else \'{n}\fi{}ski}},\ }\bibfield  {title} {\bibinfo {title} {Phase estimation without a priori phase knowledge in the presence of loss},\ }\href@noop {} {\bibfield  {journal} {\bibinfo  {journal} {Phys. Rev. A}\ }\textbf {\bibinfo {volume} {82}},\ \bibinfo {pages} {053804} (\bibinfo {year} {2010})}\BibitemShut {NoStop}%
\bibitem [{\citenamefont {Demkowicz-Dobrza{\'n}ski}\ \emph {et~al.}(2012)\citenamefont {Demkowicz-Dobrza{\'n}ski}, \citenamefont {Ko{\l}ody{\'n}ski},\ and\ \citenamefont {Gu{\c{t}}{\u{a}}}}]{demkowicz2012elusive}%
  \BibitemOpen
  \bibfield  {author} {\bibinfo {author} {\bibfnamefont {R.}~\bibnamefont {Demkowicz-Dobrza{\'n}ski}}, \bibinfo {author} {\bibfnamefont {J.}~\bibnamefont {Ko{\l}ody{\'n}ski}},\ and\ \bibinfo {author} {\bibfnamefont {M.}~\bibnamefont {Gu{\c{t}}{\u{a}}}},\ }\bibfield  {title} {\bibinfo {title} {The elusive heisenberg limit in quantum-enhanced metrology},\ }\href@noop {} {\bibfield  {journal} {\bibinfo  {journal} {Nature communications}\ }\textbf {\bibinfo {volume} {3}},\ \bibinfo {pages} {1} (\bibinfo {year} {2012})}\BibitemShut {NoStop}%
\bibitem [{\citenamefont {Mart{\'\i}nez-Vargas}\ \emph {et~al.}(2017)\citenamefont {Mart{\'\i}nez-Vargas}, \citenamefont {Pineda}, \citenamefont {Leyvraz},\ and\ \citenamefont {Barberis-Blostein}}]{martinez2017quantum}%
  \BibitemOpen
  \bibfield  {author} {\bibinfo {author} {\bibfnamefont {E.}~\bibnamefont {Mart{\'\i}nez-Vargas}}, \bibinfo {author} {\bibfnamefont {C.}~\bibnamefont {Pineda}}, \bibinfo {author} {\bibfnamefont {F.}~\bibnamefont {Leyvraz}},\ and\ \bibinfo {author} {\bibfnamefont {P.}~\bibnamefont {Barberis-Blostein}},\ }\bibfield  {title} {\bibinfo {title} {Quantum estimation of unknown parameters},\ }\href@noop {} {\bibfield  {journal} {\bibinfo  {journal} {Physical Review A}\ }\textbf {\bibinfo {volume} {95}},\ \bibinfo {pages} {012136} (\bibinfo {year} {2017})}\BibitemShut {NoStop}%
\bibitem [{\citenamefont {Mukhopadhyay}\ \emph {et~al.}(2025)\citenamefont {Mukhopadhyay}, \citenamefont {Montenegro},\ and\ \citenamefont {Bayat}}]{mukhopadhyay2025current}%
  \BibitemOpen
  \bibfield  {author} {\bibinfo {author} {\bibfnamefont {C.}~\bibnamefont {Mukhopadhyay}}, \bibinfo {author} {\bibfnamefont {V.}~\bibnamefont {Montenegro}},\ and\ \bibinfo {author} {\bibfnamefont {A.}~\bibnamefont {Bayat}},\ }\bibfield  {title} {\bibinfo {title} {Current trends in global quantum metrology},\ }\href {http://iopscience.iop.org/article/10.1088/1751-8121/adb112} {\bibfield  {journal} {\bibinfo  {journal} {Journal of Physics A: Mathematical and Theoretical}\ }\textbf {\bibinfo {volume} {58}},\ \bibinfo {pages} {063001} (\bibinfo {year} {2025})}\BibitemShut {NoStop}%
\bibitem [{\citenamefont {Cenni}\ \emph {et~al.}(2022)\citenamefont {Cenni}, \citenamefont {Lami}, \citenamefont {Acin},\ and\ \citenamefont {Mehboudi}}]{cenni2022thermometry}%
  \BibitemOpen
  \bibfield  {author} {\bibinfo {author} {\bibfnamefont {M.~F.}\ \bibnamefont {Cenni}}, \bibinfo {author} {\bibfnamefont {L.}~\bibnamefont {Lami}}, \bibinfo {author} {\bibfnamefont {A.}~\bibnamefont {Acin}},\ and\ \bibinfo {author} {\bibfnamefont {M.}~\bibnamefont {Mehboudi}},\ }\bibfield  {title} {\bibinfo {title} {Thermometry of gaussian quantum systems using gaussian measurements},\ }\href@noop {} {\bibfield  {journal} {\bibinfo  {journal} {Quantum}\ }\textbf {\bibinfo {volume} {6}},\ \bibinfo {pages} {743} (\bibinfo {year} {2022})}\BibitemShut {NoStop}%
\bibitem [{\citenamefont {Genoni}\ \emph {et~al.}(2014)\citenamefont {Genoni}, \citenamefont {Mancini},\ and\ \citenamefont {Serafini}}]{genoni2014general}%
  \BibitemOpen
  \bibfield  {author} {\bibinfo {author} {\bibfnamefont {M.~G.}\ \bibnamefont {Genoni}}, \bibinfo {author} {\bibfnamefont {S.}~\bibnamefont {Mancini}},\ and\ \bibinfo {author} {\bibfnamefont {A.}~\bibnamefont {Serafini}},\ }\bibfield  {title} {\bibinfo {title} {General-dyne unravelling of a thermal master equation},\ }\href@noop {} {\bibfield  {journal} {\bibinfo  {journal} {Russian Journal of Mathematical Physics}\ }\textbf {\bibinfo {volume} {21}},\ \bibinfo {pages} {329} (\bibinfo {year} {2014})}\BibitemShut {NoStop}%
\bibitem [{\citenamefont {Yu}\ \emph {et~al.}(2025)\citenamefont {Yu}, \citenamefont {Liu}, \citenamefont {Xue}, \citenamefont {Yang}, \citenamefont {Wang}, \citenamefont {Zhang}, \citenamefont {Cui}, \citenamefont {Yang}, \citenamefont {Li}, \citenamefont {Han} \emph {et~al.}}]{yu2025experimental}%
  \BibitemOpen
  \bibfield  {author} {\bibinfo {author} {\bibfnamefont {Y.}~\bibnamefont {Yu}}, \bibinfo {author} {\bibfnamefont {R.}~\bibnamefont {Liu}}, \bibinfo {author} {\bibfnamefont {G.}~\bibnamefont {Xue}}, \bibinfo {author} {\bibfnamefont {C.}~\bibnamefont {Yang}}, \bibinfo {author} {\bibfnamefont {C.}~\bibnamefont {Wang}}, \bibinfo {author} {\bibfnamefont {J.}~\bibnamefont {Zhang}}, \bibinfo {author} {\bibfnamefont {J.}~\bibnamefont {Cui}}, \bibinfo {author} {\bibfnamefont {X.}~\bibnamefont {Yang}}, \bibinfo {author} {\bibfnamefont {J.}~\bibnamefont {Li}}, \bibinfo {author} {\bibfnamefont {J.}~\bibnamefont {Han}}, \emph {et~al.},\ }\bibfield  {title} {\bibinfo {title} {Experimental realization of criticality-enhanced global quantum sensing via non-equilibrium dynamics},\ }\href@noop {} {\bibfield  {journal} {\bibinfo  {journal} {arXiv preprint arXiv:2501.04955}\ } (\bibinfo {year} {2025})}\BibitemShut {NoStop}%
\bibitem [{\citenamefont {Lieb}\ \emph {et~al.}(1961)\citenamefont {Lieb}, \citenamefont {Schultz},\ and\ \citenamefont {Mattis}}]{lieb1961two}%
  \BibitemOpen
  \bibfield  {author} {\bibinfo {author} {\bibfnamefont {E.}~\bibnamefont {Lieb}}, \bibinfo {author} {\bibfnamefont {T.}~\bibnamefont {Schultz}},\ and\ \bibinfo {author} {\bibfnamefont {D.}~\bibnamefont {Mattis}},\ }\bibfield  {title} {\bibinfo {title} {Two soluble models of an antiferromagnetic chain},\ }\href@noop {} {\bibfield  {journal} {\bibinfo  {journal} {Annals of Physics}\ }\textbf {\bibinfo {volume} {16}},\ \bibinfo {pages} {407} (\bibinfo {year} {1961})}\BibitemShut {NoStop}%
\bibitem [{\citenamefont {Suzuki}(2016)}]{suzuki2016explicit}%
  \BibitemOpen
  \bibfield  {author} {\bibinfo {author} {\bibfnamefont {J.}~\bibnamefont {Suzuki}},\ }\bibfield  {title} {\bibinfo {title} {Explicit formula for the holevo bound for two-parameter qubit-state estimation problem},\ }\href@noop {} {\bibfield  {journal} {\bibinfo  {journal} {Journal of Mathematical Physics}\ }\textbf {\bibinfo {volume} {57}} (\bibinfo {year} {2016})}\BibitemShut {NoStop}%
\bibitem [{\citenamefont {Gessner}\ \emph {et~al.}(2018)\citenamefont {Gessner}, \citenamefont {Pezz{\`e}},\ and\ \citenamefont {Smerzi}}]{gessner2018sensitivity}%
  \BibitemOpen
  \bibfield  {author} {\bibinfo {author} {\bibfnamefont {M.}~\bibnamefont {Gessner}}, \bibinfo {author} {\bibfnamefont {L.}~\bibnamefont {Pezz{\`e}}},\ and\ \bibinfo {author} {\bibfnamefont {A.}~\bibnamefont {Smerzi}},\ }\bibfield  {title} {\bibinfo {title} {Sensitivity bounds for multiparameter quantum metrology},\ }\href@noop {} {\bibfield  {journal} {\bibinfo  {journal} {Physical review letters}\ }\textbf {\bibinfo {volume} {121}},\ \bibinfo {pages} {130503} (\bibinfo {year} {2018})}\BibitemShut {NoStop}%
\bibitem [{\citenamefont {Albarelli}\ \emph {et~al.}(2019)\citenamefont {Albarelli}, \citenamefont {Friel},\ and\ \citenamefont {Datta}}]{albarelli2019evaluating}%
  \BibitemOpen
  \bibfield  {author} {\bibinfo {author} {\bibfnamefont {F.}~\bibnamefont {Albarelli}}, \bibinfo {author} {\bibfnamefont {J.~F.}\ \bibnamefont {Friel}},\ and\ \bibinfo {author} {\bibfnamefont {A.}~\bibnamefont {Datta}},\ }\bibfield  {title} {\bibinfo {title} {Evaluating the holevo cram{\'e}r-rao bound for multiparameter quantum metrology},\ }\href@noop {} {\bibfield  {journal} {\bibinfo  {journal} {Physical review letters}\ }\textbf {\bibinfo {volume} {123}},\ \bibinfo {pages} {200503} (\bibinfo {year} {2019})}\BibitemShut {NoStop}%
\bibitem [{\citenamefont {Razavian}\ \emph {et~al.}(2020)\citenamefont {Razavian}, \citenamefont {Paris},\ and\ \citenamefont {Genoni}}]{razavian2020quantumness}%
  \BibitemOpen
  \bibfield  {author} {\bibinfo {author} {\bibfnamefont {S.}~\bibnamefont {Razavian}}, \bibinfo {author} {\bibfnamefont {M.~G.~A.}\ \bibnamefont {Paris}},\ and\ \bibinfo {author} {\bibfnamefont {M.~G.}\ \bibnamefont {Genoni}},\ }\bibfield  {title} {\bibinfo {title} {On the quantumness of multiparameter estimation problems for qubit systems},\ }\href@noop {} {\bibfield  {journal} {\bibinfo  {journal} {Entropy}\ }\textbf {\bibinfo {volume} {22}},\ \bibinfo {pages} {1197} (\bibinfo {year} {2020})}\BibitemShut {NoStop}%
\bibitem [{\citenamefont {Yang}\ \emph {et~al.}(2025)\citenamefont {Yang}, \citenamefont {Montenegro},\ and\ \citenamefont {Bayat}}]{yang2025overcoming}%
  \BibitemOpen
  \bibfield  {author} {\bibinfo {author} {\bibfnamefont {Y.}~\bibnamefont {Yang}}, \bibinfo {author} {\bibfnamefont {V.}~\bibnamefont {Montenegro}},\ and\ \bibinfo {author} {\bibfnamefont {A.}~\bibnamefont {Bayat}},\ }\bibfield  {title} {\bibinfo {title} {Overcoming quantum metrology singularity through sequential measurements},\ }\href@noop {} {\bibfield  {journal} {\bibinfo  {journal} {arXiv preprint arXiv:2501.02784}\ } (\bibinfo {year} {2025})}\BibitemShut {NoStop}%
\bibitem [{\citenamefont {Palmieri}\ \emph {et~al.}(2021)\citenamefont {Palmieri}, \citenamefont {Bianchi}, \citenamefont {Paris},\ and\ \citenamefont {Benedetti}}]{Benedetti21}%
  \BibitemOpen
  \bibfield  {author} {\bibinfo {author} {\bibfnamefont {A.~M.}\ \bibnamefont {Palmieri}}, \bibinfo {author} {\bibfnamefont {F.}~\bibnamefont {Bianchi}}, \bibinfo {author} {\bibfnamefont {M.~G.~A.}\ \bibnamefont {Paris}},\ and\ \bibinfo {author} {\bibfnamefont {C.}~\bibnamefont {Benedetti}},\ }\bibfield  {title} {\bibinfo {title} {Multiclass classification of dephasing channels},\ }\href {https://doi.org/10.1103/PhysRevA.104.052412} {\bibfield  {journal} {\bibinfo  {journal} {Phys. Rev. A}\ }\textbf {\bibinfo {volume} {104}},\ \bibinfo {pages} {052412} (\bibinfo {year} {2021})}\BibitemShut {NoStop}%
\bibitem [{\citenamefont {Belliardo}\ \emph {et~al.}(2024)\citenamefont {Belliardo}, \citenamefont {Zoratti},\ and\ \citenamefont {Giovannetti}}]{belliardo2024applications}%
  \BibitemOpen
  \bibfield  {author} {\bibinfo {author} {\bibfnamefont {F.}~\bibnamefont {Belliardo}}, \bibinfo {author} {\bibfnamefont {F.}~\bibnamefont {Zoratti}},\ and\ \bibinfo {author} {\bibfnamefont {V.}~\bibnamefont {Giovannetti}},\ }\bibfield  {title} {\bibinfo {title} {Applications of model-aware reinforcement learning in bayesian quantum metrology},\ }\href@noop {} {\bibfield  {journal} {\bibinfo  {journal} {Physical Review A}\ }\textbf {\bibinfo {volume} {109}},\ \bibinfo {pages} {062609} (\bibinfo {year} {2024})}\BibitemShut {NoStop}%
\bibitem [{\citenamefont {Xu}\ \emph {et~al.}(2025)\citenamefont {Xu}, \citenamefont {Xiao}, \citenamefont {Huang}, \citenamefont {He}, \citenamefont {Fan},\ and\ \citenamefont {Zeng}}]{xu2025toward}%
  \BibitemOpen
  \bibfield  {author} {\bibinfo {author} {\bibfnamefont {H.}~\bibnamefont {Xu}}, \bibinfo {author} {\bibfnamefont {T.}~\bibnamefont {Xiao}}, \bibinfo {author} {\bibfnamefont {J.}~\bibnamefont {Huang}}, \bibinfo {author} {\bibfnamefont {M.}~\bibnamefont {He}}, \bibinfo {author} {\bibfnamefont {J.}~\bibnamefont {Fan}},\ and\ \bibinfo {author} {\bibfnamefont {G.}~\bibnamefont {Zeng}},\ }\bibfield  {title} {\bibinfo {title} {Toward heisenberg limit without critical slowing down via quantum reinforcement learning},\ }\href@noop {} {\bibfield  {journal} {\bibinfo  {journal} {Physical Review Letters}\ }\textbf {\bibinfo {volume} {134}},\ \bibinfo {pages} {120803} (\bibinfo {year} {2025})}\BibitemShut {NoStop}%
\end{thebibliography}%
\end{document}